\documentclass[journal,10pt]{IEEEtran}
\usepackage{graphicx}
\usepackage{subfigure}
\usepackage{float}
\usepackage{subfloat}
\usepackage{amsmath}
\usepackage{amssymb}
\usepackage{amsthm}
\usepackage{epstopdf}
\usepackage{cases}
\usepackage{color}
\usepackage{makecell}
\usepackage{cite}
\usepackage{algorithmic}
\usepackage[ruled]{algorithm2e}

\newtheorem{Theo}{Theorem}

\usepackage[normalem]{ulem}

\ifCLASSINFOpdf
\else
\fi
\usepackage{caption}
\usepackage{mathtools}

\begin{document}

%
\title{Code-Division OFDM Joint Communication and Sensing System for 6G Machine-type Communication}
%
%
%

\author{Xu Chen,~\IEEEmembership{Student Member,~IEEE,}
	Zhiyong Feng,~\IEEEmembership{Senior Member,~IEEE,}
	Zhiqing Wei,~\IEEEmembership{Member,~IEEE,}
	Ping Zhang,~\IEEEmembership{Fellow,~IEEE,}
	and Xin Yuan,~\IEEEmembership{Member,~IEEE}
	
	\thanks{Xu Chen, Z. Feng, and Z. Wei are with Beijing University of Posts and Telecommunications, Key Laboratory of Universal Wireless Communications, Ministry of Education, Beijing 100876, P. R. China (Email:\{chenxu96330, fengzy, weizhiqing\}@bupt.edu.cn).}
	\thanks{ Ping Zhang is with Beijing University of Posts and Telecommunications, State Key Laboratory of Networking and Switching Technology, Beijing 100876, P. R. China (Email: pzhang@bupt.edu.cn).}
	\thanks{X. Yuan is with Commonwealth Scientific and Industrial Research Organization (CSIRO), Australia (email: Xin.Yuan@data61.csiro.au).}
	\thanks{Corresponding author: Zhiyong Feng, Zhiqing Wei}
	\thanks{Copyright (c) 20xx IEEE. Personal use of this material is permitted. However, permission to use this material for any other purposes must be obtained from the IEEE by sending a request to pubs-permissions@ieee.org.}
}

%
%

\markboth{}%
{Shell \MakeLowercase{\textit{et al.}}: Bare Demo of IEEEtran.cls for IEEE Journals}
%



\maketitle

\pagestyle{empty}  
\thispagestyle{empty} 

\newcounter{mytempeqncnt}
\setcounter{mytempeqncnt}{\value{equation}}
\begin{abstract}


\color{black} The joint communication and sensing (JCS) system can provide higher spectrum efficiency and load-saving for 6G machine-type communication (MTC) applications by merging necessary communication and sensing abilities with unified spectrum and transceivers. In order to suppress the mutual interference between the communication and radar sensing signals to improve the communication reliability and radar sensing accuracy, we propose a novel code-division orthogonal frequency division multiplex (CD-OFDM) JCS MTC system, where MTC users can simultaneously and continuously conduct communication and sensing with each other. {\color{black} We propose a novel CD-OFDM JCS signal and corresponding successive-interference-cancellation (SIC) based signal processing technique that obtains code-division multiplex (CDM) gain, which is compatible with the prevalent  orthogonal frequency division multiplex (OFDM) communication system.} To model the unified JCS signal transmission and reception process, we propose a novel unified JCS channel model. Finally, the simulation and numerical results are shown to verify the feasibility of the CD-OFDM JCS MTC system {\color{black} and the error propagation performance}. We show that the CD-OFDM JCS MTC system can achieve not only more reliable  communication but also comparably robust radar sensing compared with the precedent OFDM JCS system, especially in low signal-to-interference-and-noise ratio (SINR) regime.

\end{abstract}

\begin{IEEEkeywords}
Machine type communications, joint communication and sensing, code-division OFDM, interference cancellation.
\end{IEEEkeywords}

%
\IEEEpeerreviewmaketitle

\section{Introduction}
%
%
%
%
\subsection{Background and Motivations}
{\color{black}
	
The number of Internet of Things (IoT) devices is predicted to grow three-fold, from 11 billion in 2019 to 30 billion in 2030~\cite{latva2019key,mahmood2020white}. By 2030, the 6G MTC system intends to fulfill tremendous IoT connections for intelligent precise control applications to support the future IoT, such as advanced driving automation (from L4 to L5) and accurate vehicle swarm control in the industrial scenarios~\cite{mahmood2020white}. These applications aim to conduct customized and precise control missions in mixed sensing/actuation/haptics scenarios, and require sensing accuracy of 0.1 m indoors or 1 m outdoors and communication reliability of 99.9999\% for the critical applications~\cite{mahmood2020white}. The tremendous connections and high function requirements will result in great spectrum congestion, and demand high spectrum and energy utilization efficiency~\cite{Liuxin2020,LiuxinIoT2019,2020Chen,2021Young}, which can not be fully addressed in 5G cellular MTC scenarios~\cite{mahmood2020six}. 

The state-of-the-art joint communication and sensing (JCS) technique is a promising technique to confront the aforementioned challenges. The JCS system can share the unified transceiver, the same spectrum and digital signal processing hardware to exploit the reflected echo of communication beam to achieve additional radar sensing function~\cite{Fangzixi2020}, and achieve immediate improvement in both the spectrum and energy efficiency, compared with the conventional communication system where the echo signal is not well used for sensing~\cite{liu2020joint}. The JCS system also benefits from mutual sharing of sensing information for improved reliability performance, e.g., using the range and Doppler knowledge to assist beamforming and channel prediction~\cite{Zhang2019JCRS}.}

\subsection{Related Works}
{\color{black} The early researches about JCS systems concentrated on the JCS waveform design based on MIMO and beamforming techniques used in communications~\cite{Fangzixi2020}. In \cite{Strum2009OFDMradar, Sturm2011Waveform}, Sturm \textit{et al.} proposed an orthogonal frequency division multiplexing (OFDM) symbol-based JCS signal processing method, which overcomes the typical drawbacks of correlation-based radar signal processing and satisfies both the radar ranging and communication requirements. In \cite{moghaddasi2016multifunctional}, the authors proposed a reconfigurable and unified multifunctional receiver for data-fusion services of radar sensing and radio communication based on time-division platform. In each time slot assigned for radar sensing or radio communication modes, the system can achieve localization function or data communication, respectively. 

Other researches on the JCS system focused on designing the JCS transmitting and receiving system architectures to exploit time and spectrum resources effectively.
Kumari \textit{et al.} \cite{Kumari2018WIFI} proposed an IEEE 802.11ad-based OFDM JCS vehicle-to-vehicle (V2V) system exploiting the preamble of a single-carrier physical layer frame to achieve V2V communication and full-duplex radar at the 60 GHz band, which benefits from the great development of the sufficient isolation and self-interference cancellation. In~\cite{Zengyonghong2018}, the authors proposed an OFDM JCS system based on the IEEE 802.11 standards for range and speed detection, which can achieve full-duplex operation through canceling the self-interference by estimating the self-interference channel between the transmitting and receiving antennas.
In \cite{Zhang2019JCRS}, Zhang \textit{et al.} proposed a practical OFDM time-division-duplex (TDD) multi-beam scheme to achieve JCS, which complies with the prevalent terrestrial packet communication system. This work utilizes multi-beam forming scheme to generate several orthogonal beams, and assigns to each beam either communication or sensing function. Based on the evolved mobile broadband (eMBB) scenario, Liu \textit{et al.} further proposed the concept of JCS base station that operates in mmWave band with TDD protocol adopting single-array transceivers \cite{liu2020joint}. The authors considered that the targets of interests can also be scatterers, and proposed the concept of successive interference cancellation (SIC) based method to suppress the mutual interference between communication and radar sensing signal reception.

However, there are some challenges that the aforementioned state-of-the-art work cannot completely handle. The features of massive and dense users of 6G MTC scenarios lead to close range between MTC users and severe mutual interference~\cite{latva2019key}. The close range problem makes the single-array JCS transceiver designed for eMBB users vulnerable to the minimum range problem, i.e., if the target of interest is too close to the JCS user, then the target may be ignored because the transceiver may not be in receiving mode when radar echo returns, which is disastrous for mobile MTC applications~\cite{2014IBFD}. In \cite{Zhang2019JCRS}, Zhang \textit{et al.} adopted double-array transceivers, which can make the receiver work constantly and thus handle the minimum range problem. The severe mutual interference problem makes the conventional OFDM JCS system hard to ensure the high-reliability requirements in low signal-to-interference-and-noise ratio (SINR) regime~\cite{mahmood2020white}. 
}

\subsection{Our Contributions}
In this paper, we propose a code-division OFDM (CD-OFDM) JCS system, which can conduct communication and radar sensing simultaneously and constantly with the unified spectrum and transceivers. Here, we consider the MTC scenarios of low speed movement. To suppress the mutual interference between communication and radar echo signals, we propose a novel CD-OFDM JCS signal and corresponding SIC based processing method, which is compatible with OFDM JCS signal processing.  In order to overcome the aforementioned problem of minimum detection range, we adopt double-array transceiver design to make the JCS MTC devices able to conduct constant radar sensing and communication simultaneously. 
%
%

The main contributions of this paper are summarized as follows. 
\begin{itemize}
	\item[1.] We propose a novel CD-OFDM JCS MTC system that is compatible with both CD-OFDM and OFDM JCS signal processing. This system can achieve radar sensing and communication between MTC devices simultaneously and constantly with the unified spectrum and transceivers. 
	
	\item[2.] {\color{black} We propose a novel SIC-based CD-OFDM and OFDM JCS signal processing method for the CD-OFDM JCS MTC system that can achieve the code-division gain to suppress the  multiple access interference for communication, including the radar echo signal and co-channel interference of other radio users. }
	
	\item[3.] {\color{black} We propose a novel unified JCS channel model based on the MIMO communication and radar channel models, presenting comprehensive quantitative relation among the fading coefficients, average power, delay and Doppler of communication and radar echo channels.}
\end{itemize}
\subsection{Outline of This Paper}
The remaining parts of this paper are organized as follows. 
In Section \ref{sec:system-model}, we describe the CD-OFDM JCS MTC system model and propose the JCS MIMO channel model. 
Section \ref{sec: Communication} proposes the SIC-based CD-OFDM JCS signal processing method.
In section \ref{sec:Numerical-results}, the simulation results are presented. 
Section \ref{sec:conclusion} concludes this paper.

Without particular claim, we adopt the following notations. Bold uppercase letters denote matrices (i.e., $\textbf{M}$); bold lowercase letters denote column vectors (i.e., $\textbf{v}$); scalers are denoted by normal font (i.e., $\gamma$); the entries of vectors or matrices are referred to with parenthesis, for instance, the $q$th entry of vector $\textbf{v}$ is $\textbf{v}(q)$, and the entry of the matrix $\textbf{M}$ at the $m$th row and $q$th column is $\textbf{M}(m,q)$; $\textbf{I}_{Q}$ is the identity matrix with dimension $Q \times Q$; matrix superscripts $\left(\cdot\right)^H$, $\left(\cdot\right)^{*}$ and $\left(\cdot\right)^T$ denote Hermitian transpose, complex conjugate and transpose, respectively; $\rm Re\left( \cdot \right)$ and $\rm Im \left( \cdot \right)$ are the real and imaginary parts of complex number, $E\left( \cdot \right)$ represents the expectation of random variable, $\left\lfloor\cdot \right\rfloor$ denotes the floor function, and $\delta\left(t-t_0\right)$ is the unit pulse function, Besides, we use $\left(\cdot\right)^{-1}$ to denote inverse of matrix, $\left(\cdot\right)^{\dag}$ to denote the pseudo-inverse of the matrix, and $diag\left(\textbf{v}\right)$ to denote a diagonal matrix with the entries of $\textbf{v}$ on the diagonal.

\section{System Model}\label{sec:system-model}

\subsection{The CD-OFDM JCS MTC System Model} \label{subsec:Duplex model}
\begin{figure}[!t]
	\centering
	\includegraphics[width=0.37\textheight]{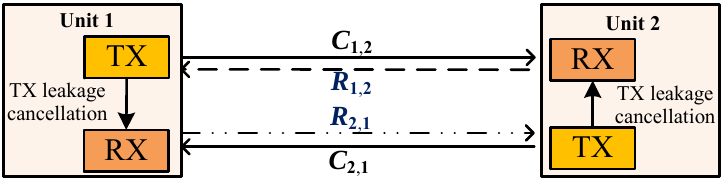}%
	\DeclareGraphicsExtensions.
	\caption{The JCS system model. MTC users 1 and 2 conduct simultaneous bi-directional communication and sensing with each other through LoS links. A leakage cancellation module between TxA and RxA is used to cancel the self-interference.}
	\label{fig:Duplex model}
\end{figure}

We consider a CD-OFDM JCS MTC system of low speed movement, where MTC devices conduct simultaneous and constant radar sensing and communication with the unified spectrum and transceiver\footnote{In this paper, all MTC devices are machines with considerable radar cross section (RCS), such as low-speed vehicles on the road and unmanned aerial vehicles. }. As illustrated in Fig.~\ref{fig:Duplex model}, MTC users 1 and 2  conduct simultaneous bi-directional communication and radar sensing through line-of-sight (LoS) links. Each user is equipped with a double-array JCS transceiver composed of a transmitting array (TxA) and a receiving array (RxA) to generate the transmitting beams (TxBs) and receiving beams (RxBs), respectively. {\color{black} We assume that the transmit power for users 1 and 2 on each subcarrier is $P_1$ and $P_2$, respectively, and the antenna arrays are all uniform linear arrays (ULAs).} The isotropic antenna numbers of TxA and RxA are denoted by $M$ and $N$, respectively. 
{\color{black} In order to solve the aforementioned minimum range problem, in-band full-duplex (IBFD) operation is required to make RxA work constantly~\cite{2014IBFD}. In this case, the self-interference from TxA is large enough to ruin the communication and echo receiving. Thus, an isolation shielding plate and leakage cancellation module between TxA and RxA are required to alleviate or even cancel the self-leakage interference\footnote{Because the leakage signal can be canceled, for simplicity, the residual self-interference caused by quantization error \cite{2014IBFD} is modeled into noise and does not appear in the equations in this paper.}\cite{2014IBFD}. Compared with the double-array transceiver design in~\cite{Zhang2019JCRS} that adopts separate spectra for communication in TDD mode and constant radar detection, respectively, our proposed CD-OFDM JCS transceiver uses the same spectrum for both communication and sensing simultaneously and can achieve IBFD JCS operation.}

In the JCS MTC system, users 1 and 2 both generate TxB and RxB pointed to each other to establish communication links (CLs), i.e., $C_{1,2}$ and $C_{2,1}$, and radar echo links (RELs), i.e., $R_{1,2}$ and $R_{2,1}$, as illustrated in Fig. 1. We assume that the perfect communication channel state information (CSI) is obtained at both users 1 and 2, i.e., the CSI matrices $\textbf{H}_{C,12}$ and $\textbf{H}_{C,21}$ for $C_{1,2}$ and $C_{2,1}$ can be well estimated, respectively, {\color{black} and the directions of arrival (DoA) is also detected for beamforming through multiple signal classification algorithm (MUSIC) or other angle detection methods}. We also assume that the communication channel reciprocity holds as the users exploit the same spectrum and conduct communication simultaneously \cite{Ding2019Channel}.

As illustrated in Fig.~\ref{fig:Duplex model}, user 1 transmits the JCS signal to user 2 through the CL $C_{1,2}$, and receives the reflected signals through the REL $R_{1,2}$. Once user 2 receives the JCS signal of user 1, user 2 transmits the JCS signal to user 1 through the CL $C_{2,1}$ and receives reflected signals through the REL $R_{2,1}$. The received signals at user 1 are composed of the communication signal from user 2 and the radar echo signal, while the received signals at user 2 consist of the communication signal from user 1 and the radar echo signal. It is noted that the radar echo signal is weaker than the communication signal because the radar echo transmits twice the distance as the communication signal. Thus, the radar echo signal can be regarded as small interference imposed on the communication signal, and  can only be obtained after the communication signal is canceled.

\begin{figure}[!t]
	\centering
	\includegraphics[width=0.5\textwidth]{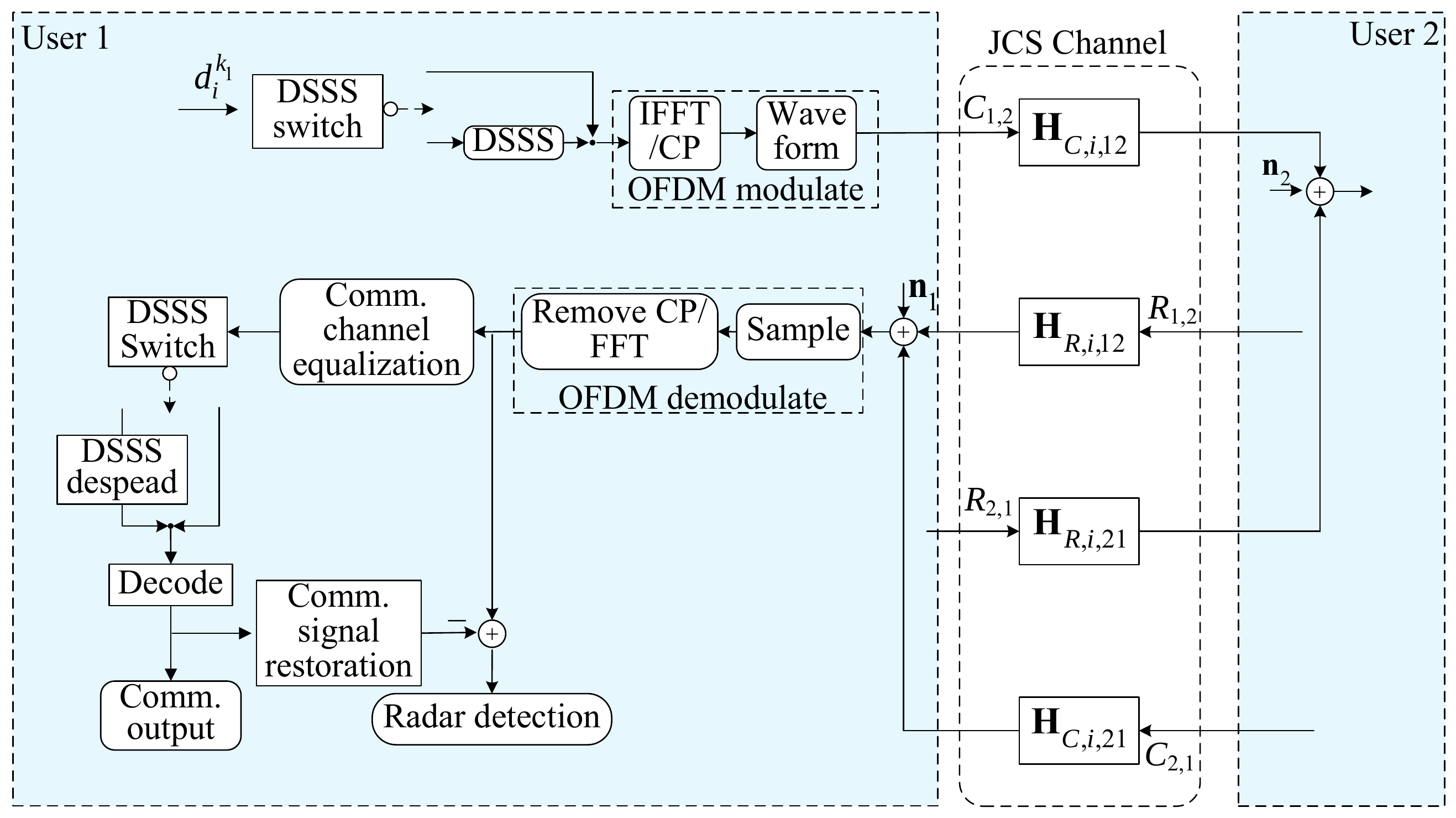}%
	\DeclareGraphicsExtensions.
	\caption{The diagram of CD-OFDM JCS signal processing. DSSS switch decides whether CD-OFDM signal or OFDM signal is used. In high SINR regime, the DSSS switch turns to the void, and only the OFDM modulator is used. Otherwise, in  low SINR regime, the switch turns to the DSSS module, and the original symbols are spread by DSSS to generate CD-OFDM signal.}
	\label{fig:JSC_signal processing}
\end{figure}
Fig.~\ref{fig:JSC_signal processing} illustrates the diagram of signal processing of the CD-OFDM JCS system. Note that user 2 has the same JCS signal processing system as user 1. An SINR threshold is assumed at the direct-sequence-spectrum-spread (DSSS) switch module. By comparing the SINR threshold with the SINR obtained in the communication channel estimation, the DSSS switch decides whether CD-OFDM signal or OFDM signal is used. {\color{black} In low SINR regime, the switch turns to the DSSS module, and then the original symbols are first spread by the DSSS code book matrix and then modulated by the OFDM modulator to generate CD-OFDM signals, which enjoys code-division-multiplex (CDM) gain to enhance processing SINR and reliability at the cost of high computation complexity. In high SINR regime, the DSSS switch turns to the void, i.e., the DSSS code book matrix is set as identity matrix, and therefore the original symbols are directly modulated by OFDM modulator, as OFDM signal can satisfy the reliability constraints in the high SINR regime without additional computation complexity.} 

{\color{black} Note that the CD-OFDM JCS system can be achieved by modifying the frequency domain symbol processing modules of the MIMO OFDM communication system, the advanced DoA detection algorithms that are used in the MIMO OFDM system, such as MUSIC algorithm, can be applied to the CD-OFDM JCS system. Therefore, the angle detection is not our key innovation interest in this paper.}

After the superposed signal is received, composed of communication, radar echo and interference-and-noise signals, it is first processed by OFDM demodulator, and then the superposed JCS symbols with interference-plus-noise are obtained. Regarding the radar echo as the small interference imposed on communication signal, communication demodulation is first conducted. DSSS switch module decides whether DSSS despreading is used or not. After decoding the communication symbols as shown in Section~\ref{sec:Communication Demodulation and Radar Signal Acquirement of CD-OFDM JCS Signal}, the communication signal can be restored and then canceled from the superposed received signal, which generates the radar signal plus noise. At last, radar detection can be conducted as shown in Section \ref{sec:the-acquisition-of-radar-echo}.

\subsection{The CD-OFDM Signal Model}\label{sec:Signal model}

The $i$th CD-OFDM symbol transmitted by the $k$th code channel in time domain is expressed as \cite{wang2010joint}

\begin{equation}\label{equ:CDMA_OFDM_T}
{s}_i^k\left( t \right) = \frac{1}{{\sqrt {{T_s}} }}\sum\limits_{m = 0}^{{N_c} - 1} {{\bf{c}}_i^k\left( m \right)d_i^k{e^{j 2\pi {{m}\Delta f (t-iT_s)}}}rect\left( {\frac{t-iT_s}{{{T_{s}}}}} \right)},
\end{equation}
where $N_c$ is the number of subcarriers, $\Delta f = 1/T = B/{N_c}$ is the subcarrier spacing with $T$ and $B$ being the duration of elementary CD-OFDM symbol and bandwidth respectively, ${{\bf{ c}}_i^k}$ is the DSSS code vector that is utilized by the $k$th code channel to transmit the $i$th CD-OFDM symbol, $m = 0,1,...,N_c-1$ is the index of the subcarrier, ${{\bf{ c}}}_i^k\left( m \right)$ is the $m$th entry of ${{\bf{ c}}_i^k}$, $d_i^k$ is the $i$th information symbol transmitted by the $k$th code channel, $T_{s} = T + T_g$ is the duration time of each CD-OFDM symbol with $T_g$ being the duration of cyclic prefix, and $rect\left( t/T_{s}\right)$ is the rectangle window function of duration $T_{s}$. The time interval of the $i$th CD-OFDM symbol is $t \in \left[ {\left( i-1 \right)T_{s}, iT_{s}} \right]$. The orthogonal Walsh-Hadamard code is adopted as the DSSS code in this paper. There are $N_c$ codes in the code book, i.e., $k=0,1,...,N_c - 1$. The orthogonality among ${{\bf{ c}}}_i^k$ can be given by \cite{Blel2015Walsh}
\begin{equation}
{\left( {{{\bf{ c}}}_i^k} \right)^H}{{\bf{ c}}}_i^l = \left\{ \begin{array}{l}
1,\quad{{ for \ k = l}}\\
0,{{ \quad otherwise}}
\end{array} \right..
\end{equation}

The discrete time-domain signal obtained by sampling \eqref{equ:CDMA_OFDM_T} with sampling rate $N_c/T$ within the interval $[(i-1)T_s + T_g,iT_s]$ is expressed as \cite{wang2010joint}
\begin{equation}\label{equ:CDMA_OFDM_SIGNAL}
{{\bf{ s}}}_i^k\left( n \right) = \frac{1}{{\sqrt N_c }}d_i^k\sum\limits_{m = 0}^{N_c - 1} {{{\bf{c}}}_i^k\left( m \right){e^{j{2\pi \frac{m}{N_c}n}}}} rect(\frac{n}{N_c}),
\end{equation}
where $rect(\cdot)$ is the discrete rectangle window function, and $n = {0, 1,..., N_c- 1}$. The signal vector of the $i$th CD-OFDM symbol transmitted with the $k$th code channel is
\begin{equation}
{{\bf{ s}}}_i^k = {\left[ {{\bf{ s}}_i^k\left(0 \right),{\bf{ s}}_i^k\left( 1 \right),...,{\bf{ s}}_i^k\left(N_c- 1 \right)} \right]^T}.
\end{equation}

We can rewrite \eqref{equ:CDMA_OFDM_SIGNAL} with Fourier transform regardless of the rectangle window function, that is 
\begin{equation}
{{\bf{ s}}}_i^k = d_i^k \cdot IFFT\left( {{{\bf{ c}}}_i^k} \right){\rm{ = }}d_i^k \cdot {{\bf{F}}^{ - 1}}{{\bf{ c}}}_i^k = d_i^k \cdot {{\bf{ \bar{c}}}}_i^k,
\end{equation}
where $IFFT(\cdot)$ is the inverse fast Fourier transform, and ${{\bf{ \bar{c}}}}_i^k = {{\bf{F}}^{ - 1}}{{\bf{ c}}}_i^k$ is the inverse fast Fourier transform (IFFT) of ${{\bf{ c}}}_i^k$. Moreover, ${\bf{F}}$ denotes the fast Fourier transform (FFT) matrix and is given by
\begin{equation} \label{equ:F}
{\bf{F}} = \frac{1}{{\sqrt N_c }}\left( {\begin{array}{*{20}{c}}
	1&1& \cdots &1\\
	1&W& \cdots &{{W^{\left( {{N_c} - 1} \right)}}}\\
	1& \vdots & \cdots & \vdots \\
	1&{{W^{\left( {{N_c} - 1} \right)}}}& \cdots &{{W^{{{\left( {{N_c} - 1} \right)}^2}}}}
	\end{array}} \right),
\end{equation}
where $W = {e^{ - j\frac{{2\pi }}{N_c}}}$. The superposed time-domain signal vector of $NC$ used code channels is 
\begin{equation} \label{equ:Si_matrix}
{{{\bf{ s}}}_i} = \sum\limits_{k = 0}^{NC - 1} {d_i^k{{\bf{\bar{c}}}}_i^k}.
\end{equation}

The frequency-domain signal vector corresponding to \eqref{equ:Si_matrix} is
\begin{equation} \label{equ:Si_matrix_fre}
{{\bar{ \bf{ s}}}_i} = {\bf{F}}{{{\bf{ s}}}_i} = \sum\limits_{k = 0}^{NC - 1} {d_i^k{{\bf{{c}}}}_i^k}={{\bf{C}}_i {\bf{d}}_i}
,
\end{equation}
where ${\bf{d}}_i = \left[ {d_i^{{0}},d_i^{{1}},...,d_i^{{NC-1}}} \right]^T$ is the information symbol vector with the $k$th entry being the information symbol transmitted by the $k$th code channel, ${\bf{C}}_i=\left[ {\bf{c}}_i^{0},{\bf{c}}_i^{1},...,{\bf{c}}_i^{NC-1} \right]$ is the {\color{black} code book }matrix stacked by $NC$ DSSS codes. {\color{black} We assume that all the subcarrier symbols are independent with the same power $P_s$, i.e., $E\left[d_i^{{j_1}}{\left( {d_i^{{j_2}}} \right)^*} \right] = 0$, for $j_1 \ne j_2$, otherwise, $E\left[d_i^{{j_1}}{\left( {i^{{j_2}}} \right)^*} \right] = P_s$, then the power of each subcarrier symbol of ${{\bar{ \bf{ s}}}_i}$ is $NC \times P_s$. Therefore, $\frac{{{\bar{ \bf{ s}}}_i}}{\sqrt{NC}}$ is the normalized signal vector where all the entries have the same power as the entries of ${\bf{d}}_i$. The CD-OFDM frequency domain symbols are the linear superposition of several information symbols that are spread to a certain number of subcarriers by the code book matrix. By exploiting the orthogonality of the code book, the CDM gain can be obtained.}

%

\begin{figure}[!t]
	\centering
	\includegraphics[width=0.35\textheight]{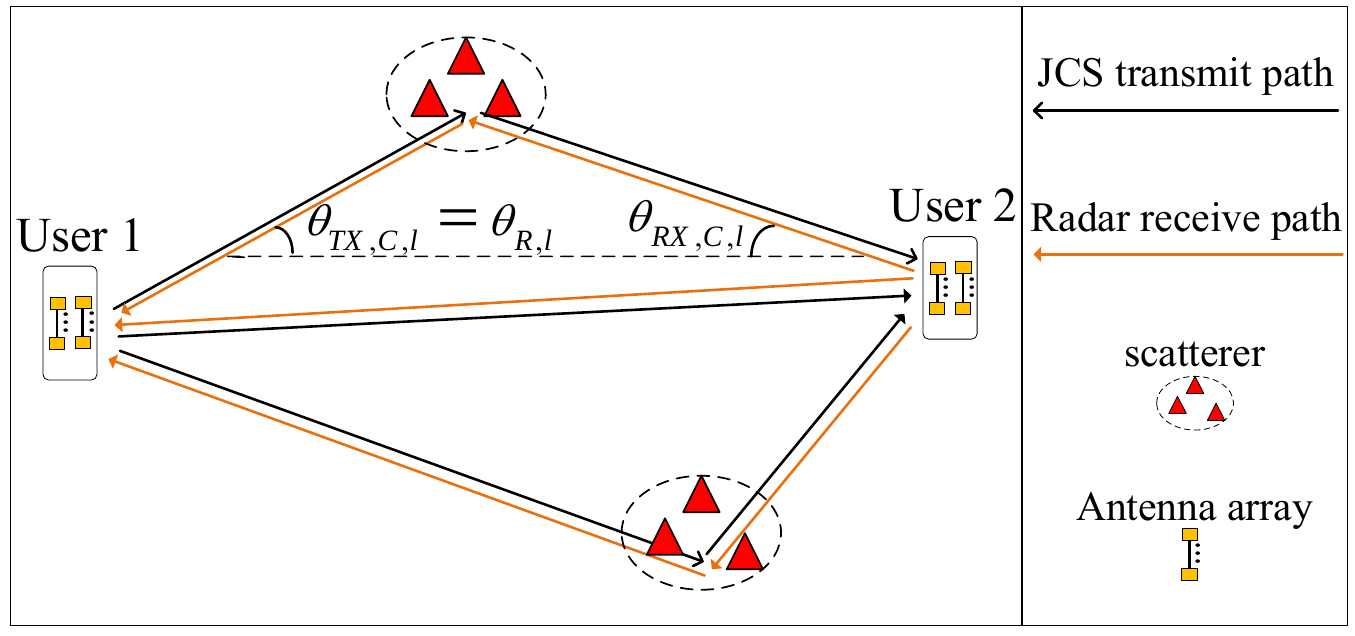}%
	\DeclareGraphicsExtensions.
	\caption{The unified JCS channel model. Users 1 and 2 conduct communication and radar sensing with each other in the LoS path with several non-LoS multipaths.}
	\label{fig:JSC_channel}
\end{figure}

\begin{figure*}[!ht]
	\normalsize
	\setcounter{equation}{17}
	\begin{equation}\label{equ:Communication_channel_fre}
		\bar h_{C,21,i}(m \Delta f) = {{\bf{ w}}_{RX,C}^H}\left( {\sum\limits_{l = 0}^{L - 1} {{b_{C,l}}{e^{j2\pi {f_{d,c,l}}(\tau_{C,l}+i{T_s})}} {e^{ - j2\pi {m \Delta f}\tau_{C,l}}} {\bf{ a}}_N\left( {{\theta _{RX,C,l}}} \right){{{\bf{ a}}}_M^T}\left( {{\theta _{TX,C,l}}} \right)} } \right){{\bf{ w}}^*_{TX}},
	\end{equation}
	\hrulefill 
\end{figure*}

\begin{figure*}[!ht]
	\normalsize
	\begin{equation}\label{equ:Radar_channel_Tx_fre} 
		{\bar h_{R,12,i}}(m \Delta f) = {{\bf{w}}_{RX,R}^H} \left( \sum\limits_{l = 0}^{L - 1} {b_{R,l}}{e^{j2\pi {f_{d,R,l}}(\tau_{R,l}+iT_s)}} {e^{ - j2\pi {m \Delta f}\tau_{R,l}}} {\bf{a}}_N\left( {{\theta _{R,l}}} \right) {{{\bf{a}}}_M^T}\left( {{\theta _{R,l}}} \right) \right)  {{\bf{w}}^*_{TX}},
	\end{equation}
	\hrulefill 
\end{figure*}

\subsection{The Unified JCS Channel Model}
{\color{black}
{\color{black} 
The JCS channel model is basic for modeling the transmitting and receiving process of JCS signal. MIMO communication and radar channel models are correlated and similar in forms\cite{Fisher2004radar,Zhang2019JCRS}. As illustrated in Fig.~\ref{fig:JSC_channel}, the communication channel is a one-direction transmitting channel, while the radar channel is a dual-direction channel consisting of both transmitting and reflecting channels. Moreover, there is a strong LoS path between user 1 and user 2 in the JCS applications. The delay of radar channel is double of the communication channel, thus, the average power and phase change of the JCS communication and radar channels are also correlated considering that the radar cross section (RCS) contributes additional fading. 

However, to the best of our knowledge, there is still no such  JCS MIMO channel models that explicitly show the comprehensive quantitative relation among the fading coefficients, average power, delay and Doppler in detailed expressions, which is required to better model the JCS propagation process. In this section, we intend to propose the JCS MIMO channel model that can handle this problem. 
}

The steering vectors of ULAs are expressed as follows:
\setcounter{equation}{8}
\begin{equation}
{{\bf{a}}_M{({\theta})}} = {\left[ {1,{e^{ j\frac{{2\pi }}{\lambda }d\sin {\theta}}},...,{e^{ j\frac{{2\pi }}{\lambda }\left( {M - 1} \right) d\sin {\theta}}}} \right]^T},
\end{equation}
where $\theta$ is the angle of arrival (AoA) or angle of departure (AoD), $\lambda$ is the wavelength of carrier, $d$ is the antenna elements spacing, and $M$ is the number of antenna elements. Without loss of generality, we set $d = \lambda/2$.

\setcounter{mytempeqncnt}{\value{equation}}
{\color{black}
The communication and radar channel responses of $C_{1,2}$ and $R_{1,2}$ for the $i$th CD-OFDM or OFDM symbol can be respectively given by \cite{Zhang2019JCRS}
\begin{equation} \label{equ:Commu_channel_T_beam}
{h_{C,12,i}}\left( t \right) = {{\bf{w}}_{RX,C}^H} \left({{\textbf{H}}_{C,12,i}}(t)\right) {{\bf{w}}^*_{TX}},
\end{equation}
and
\begin{equation} \label{equ:Radar_channel} 
{h_{R,12,i}}\left( t \right) = {{\bf{w}}_{RX,R}^H}\left( {{{\bf{H}}_{R,12,i}}\left( t \right)} \right){{\bf{w}}^*_{TX}},
\end{equation}
where ${{\bf{w}}_{TX}}$ is the normalized JCS transmitting beamforming vector that is shared by communication and radar channels, and ${{\bf{w}}_{RX,C}}$ and ${{\bf{w}}_{RX,R}}$ are the normalized communication and radar receiving beamforming vector, respectively. Moreover, ${{\textbf{H}}_{C,12,i}}(t)$ and ${{{\bf{H}}_{R,12,i}}\left( t \right)}$ are the communication and radar response matrices, respectively, which can be expressed as
\begin{equation} \label{equ:Commu_channel_T}
\begin{aligned}
{{\textbf{H}}_{C,12,i}}\left( t \right) \!=\! &\sum\limits_{l = 0}^{L - 1} \!{{b_{C,l}}{e^{j2\pi {f_{d,C,l}}(t+iT_s)}}{\bf{a}}_N \! \left( {{\theta _{RX,C,l}}} \right){{{\bf{a}}}_M^T}\!\left( {\theta _{TX,C,l}} \right)} \\
&\times\!\delta \left( {t - {\tau _{C,l}}} \right),
\end{aligned}
\end{equation}
and
\begin{equation} \label{equ:Radar_channel_mat} 
\begin{aligned}
{{{\bf{H}}_{R,12,i}}\left( t \right)} = &\sum\limits_{l = 0}^{L - 1} {b_{R,l}}{e^{j2\pi {f_{d,R,l}}(t+iT_s)}} {\bf{a}}_N\left( {{\theta _{R,l}}} \right) {{{\bf{a}}}_M^T}\left( {{\theta _{R,l}}} \right)\\
&\times \delta \left( {t - {\tau _{R,l}} } \right),
\end{aligned}
\end{equation}
respectively, where $l$ is the index of multipath with $l=0$ being the index of the LoS path, $L$ is the number of multipath, ${b_{C,l}}$ and ${b_{R,l}}$ are the complex-valued fading factors for the $l$th communication and radar multipaths, respectively, with ${b_{C,0}} = \frac{\lambda}{(4\pi R_0)} \times 1$, $b_{C,l} = \frac{\lambda}{(4\pi R_l)} \times b_l (l=1,2,...,L-1)$~\cite{channel2002}, ${b_{R,l}} = \sqrt{\frac{\lambda^2}{(4\pi)^3{R_{l}}^4}}\times{b_{r,l}}$ ($l = 0,1,...,L-1$) \cite{Strum2009OFDMradar,richards2010principles}, $b_l$ and ${b_{r,l}}$ are the small-scale fading of communication multipath and radar cross section (RCS) respectively, $R_{l}$ is the length of the $l$th multipath with $R_0$ being the length of the LoS path, $b_l$ and ${b_{r,l}}$ are the independent and identically distributed (i.i.d.) random variables that follow zero-mean complex-valued Gaussian distributions with variance $\sigma_{c,l}^2$ and $\sigma_{r,l}^2$, respectively, $\theta _{RX,C,l}$ and ${{\theta _{TX,C,l}}}$ are the communication AoA and the AoD for the $l$th multipath, respectively, ${\bf{a}}_N\left(\theta _{RX,C,l}\right)$ and ${{\bf{a}}}_M\left( \theta _{TX,C,l} \right)$ are the receiving and transmitting communication steering vectors for the $l$th multipath, respectively, $f_{d,C,l}$ and $f_{d,R,l}$ are the Doppler frequency shift for the $l$th communication and radar multipath, respectively, ${\tau_{C,l}}$ and ${\tau_{R,l}}$ are the time delay for the $l$th communication and radar multipath, respectively. Here, 
\begin{equation} \label{equ:f_d_c}
{f_{d,C,l}} = \frac{{{v_{rel}}}}{{{c_0}}}{f_c},
\end{equation}
\begin{equation} \label{equ:f_d_r}
f_{d,R,l} = \frac{{2{v_{rel}}}}{{{c_0}}}{f_c},
\end{equation}
\begin{equation} \label{equ:tau_c_l}
{\tau _{C,l}} = \frac{R_l}{c_0},
\end{equation}
and
\begin{equation} \label{equ:tau_Rl}
\tau_{R,l} = \frac{2R_l}{c_0}.
\end{equation}
where $v_{rel}$ is the radial relative velocity between MTC users 1 and 2, $c_0$ is the speed of light, and $f_c$ is the carrier frequency. 

Note that ${\theta_{TX,C,l}} = {\theta _{R,l}}$ holds because the communication and radar functions share the same transceiver and transmitting beams. Moreover, we can see that $R_l$ builds the quantitative link among $\tau _{C,l}$, $\tau _{R,l}$, $b_{C,l}$ and $b_{R,l}$, which reveals that the large-scale power fading of the JCS communication channel is proportional to that of the radar channel with factor $4 \pi R_l^2$ because the propagation distance and delay of the radar signal are twice of that of the communication signal. The reflection channel fading and the RCS are the main different fading factors between the JCS communication and radar channels. The channel response matrix of $C_{2,1}$, ${{\textbf{H}}_{C,21,i}}\left( t \right)$, is equal to the transpose of ${{\textbf{H}}_{C,12,i}}\left( t \right)$ because of the channel reciprocity~\cite{Ding2019Channel}. Thus, ${h_{C,12,i}}\left( t \right) = {h_{C,21,i}}\left( t \right)$ holds, where ${h_{C,21,i}}\left( t \right)$ is the channel response of $C_{2,1}$ in the form of \eqref{equ:Commu_channel_T_beam}. 

By substituting \eqref{equ:Commu_channel_T} and \eqref{equ:Radar_channel_mat} into  \eqref{equ:Commu_channel_T_beam} and \eqref{equ:Radar_channel}, and then applying Fourier transform to \eqref{equ:Commu_channel_T_beam} and \eqref{equ:Radar_channel}, respectively, we can obtain the frequency-domain response of the JCS channel as shown in \eqref{equ:Communication_channel_fre} and \eqref{equ:Radar_channel_Tx_fre}. On one hand, the communication CSI in \eqref{equ:Communication_channel_fre} has to be first estimated to equalize the communication channel distortion \cite{goldsmith2005wireless}. On the other hand, the radar detection is to estimate the Doppler shift and the time delay of reflected signals by processing the radar channel information in \eqref{equ:Radar_channel_Tx_fre} obtained from the correlation between the received radar signal and the known transmitting signal\cite{sturm2010performance,Fisher2004radar}.
}

\section{JCS Signal Processing}\label{sec: Communication}

The processing of the JCS CD-OFDM signal is illustrated in Fig.~\ref{fig:CDM-OFDM modulation}. Here, we take the signal processing procedure of user 1 as an illustration, and user 2 has the same processing procedures as user 1. User 1 first demodulates the communication signal of user 2, regarding the radar echo signals as the small interference. Then, the communication signal of user 2 is restored with known CSI and canceled from the superposed signals. Finally, the remaining signals are the radar signal and additive-white-Gaussian-noise (AWGN), only if the communication demodulation is not erroneous. Obviously, the bit error rate (BER) of communication demodulation and SINR both directly influence the accuracy of the JCS radar detection.

{\color{black} Compared with the conventional OFDM JCS system, the code-division processing results in CDM gain, which makes the proposed CD-OFDM JCS system can improve the reliability and can work in the low-SINR regime. Moreover, the improvement of the reliability can suppress the error propagation greatly and further ensure the IBFD radar sensing performance.}

\begin{figure*}[!t]
	\centering
	\includegraphics[width=0.75\textheight]{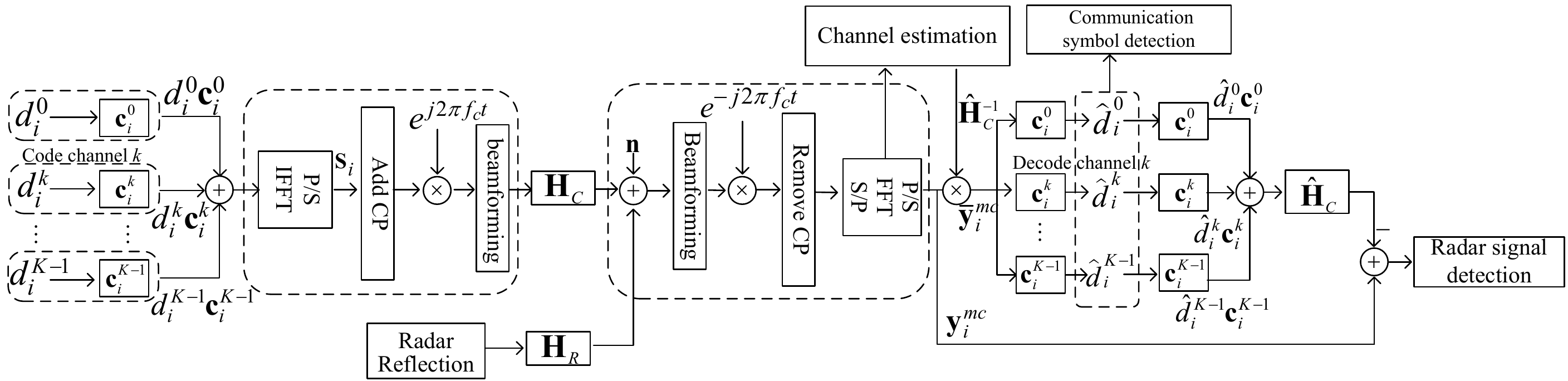}%
	\DeclareGraphicsExtensions.
	\caption{The diagram of CD-OFDM JCS signal processing.}
	\label{fig:CDM-OFDM modulation}
\end{figure*}
\subsection{Communication Demodulation and Radar Signal Acquirement}\label{sec:Communication Demodulation and Radar Signal Acquirement of CD-OFDM JCS Signal}

\setcounter{equation}{19}

{\color{black} 
	The $i$th frequency-domain CD-OFDM symbol vector received by user 1 is given by\footnote{This statement is from the view of user 1. User 2 has the opposite view to the user 1. The theoretical results of user 2 also have the same analysis procedures.} 
	\begin{equation} \label{equ:multiple channel signals}
	{\bf{y}}_{i}^{mc} = \underbrace{ {\bar{\textbf{H}}_{C,21,i}}\frac{ \sqrt{P_2} \bar{ \bf{ s}}^{k_2}_i}{{\sqrt {N{C_2}} }} }_{{\mathop{\rm communication \; signal}}} + \underbrace{ {{\bar{\textbf{H}}}_{R,12,i}}\frac{\sqrt{P_1} \bar{\bf{ s}}^{k_1}_i}{{\sqrt {N{C_1}}}} }_{{\mathop{\rm radar\;echo}}} + \; \overline {\bf{n}}_i,
	\end{equation}
	where $\overline {\bf{n}}_i$ is the complex-valued co-channel communication interference plus noise (IPN) with zero mean and covariance $\sigma_{n}^2 \textbf{I}_{N_c}$, $\bar{ \bf{ s}}^{k_2}_i$ is the JCS CD-OFDM communication symbol transmitted by MTC user 2, $\bar{ \bf{ s}}^{k_1}_i$ is the JCS CD-OFDM radar echo symbol received by MTC user 1, $\frac{1}{{\sqrt {N{C_1}}}}$ and $\frac{1}{{\sqrt {N{C_2}}}}$ are both normalization factors to ensure the unit transmit power, as demonstrated in Section~\ref{sec:Signal model}, and the radar echo is regarded as the small co-channel interference to the communication demodulation of MTC user 1. Here,
	\begin{equation} \label{equ:s_i}
	\begin{array}{l}
	\bar{ \bf{ s}}^{k_1}_i = {{{\bf{C}}_i^{{k_1}}{\bf{d}}_i^{{k_1}}}}\\
	\bar{ \bf{ s}}^{k_2}_i = {{{\bf{C}}_i^{{k_2}}{\bf{d}}_i^{{k_2}}}}
	\end{array},
	\end{equation}
	where ${\bf{d}}_i^{{k_1}}$ and ${\bf{d}}_i^{{k_2}}$ are the information symbol vectors of MTC users 1 and 2, respectively, and ${\bf{C}}_i^{{k_1}}$ and ${\bf{C}}_i^{{k_2}}$ are the DSSS code matrices of MTC users 1 and 2, respectively, both are demonstrated in Section~\ref{sec:Signal model}. 
	${{\bar{\textbf{H}}}_{C,21,i}}$ and ${{\bar{\textbf{H}}}_{R,12,i}}$ are frequency-domain JCS communication and radar channel response, respectively. Moreover, 
	\begin{equation} \label{equ:H_c_21}
	{\bar{\textbf{H}}_{C,21,i}} = diag\left(\bar{\textbf{h}}_{C,21,i}\right),
	\end{equation}
	\begin{equation} \label{equ:H_R_12}
	{\bar{\textbf{H}}_{R,12,i}} = diag\left(\bar{\textbf{h}}_{R,12,i}\right),
	\end{equation}
	where  \[ \bar{\textbf{h}}_{C,21,i}\!=\!\left[\bar h_{C,21,i}(0), \bar h_{C,21,i}(\Delta f),...,\bar h_{C,21,i}(\left(N_c \! - \! 1\right)\Delta f)\right]^{T}\!,\]
	and
	\[\bar{\textbf{h}}_{R,12,i}\!=\!\left[\bar h_{R,12,i}(0), \bar h_{R,12,i}(\Delta f),...,\bar h_{R,12,i}(\left(N_c \! - \! 1\right)\Delta f)\right]^{T}\!.\]
Left-multiply both sides of \eqref{equ:multiple channel signals} with the inverse of ${{\bar{\textbf{H}}}_{C,21,i}}$, i.e., ${{\bar{\textbf{H}}}_{C,21,i}}^{ - 1}$, and substitute \eqref{equ:s_i} into \eqref{equ:multiple channel signals}, then we have 
\begin{equation} \label{equ:multiple channel signals_after_Channel_equalization}
\overline {\bf{y}} _{i}^{mc} = \frac{\sqrt{P_2}{{\bf{C}}_i^{{k_2}}{\bf{d}}_i^{{k_2}}}}{{\sqrt {N{C_2}} }} + {{\bar{\textbf{H}}}_{C,21,i}}^{ - 1}\left( {\bar{\textbf{H}}}_{R,12,i} \frac{\sqrt{P_1}{\bf{C}}_i^{{k_1}}{\bf{d}}_i^{{k_1}}}{\sqrt{NC_1}}  + \overline {\bf{n}}_i \right).
\end{equation}

Using the Hermitian matrix of the code channel matrix ${\bf{C}}_i^{{k_2}}$ to despread \eqref{equ:multiple channel signals_after_Channel_equalization}, we obtain the despreaded communication signal as
\begin{equation} \label{equ:multiple channel signals_demodulated}
\begin{aligned}
\hat {\bf{y}} _{i}^{mc} &= \! \sqrt{\frac{NC_2}{P_2}} {\left( {{\bf{C}}_i^{{k_2}}} \right)^H}\overline {\bf{y}} _{i}^{mc} \\
&= \!{\bf{d}}_i^{{k_2}} \!\! + \!\! \underbrace{ \sqrt{\frac{NC_2}{P_2}}{\left(\!{{\bf{C}}_i^{{k_2}}} \! \right)^{\!\!H}} \!\! {{\bar{\textbf{H}}}^{ - 1}_{C,21,i}} \! \left(\!\!{{{\bar{\textbf{H}}}_{R,12,i}} \frac{\sqrt{P_1} {\bf{C}}_i^{{k_1}} {\bf{d}}_i^{{k_1}}}{\sqrt{NC_1}} \!\! + \! \overline {\bf{n}}_i } \!\right) }_{{\mathop{\rm interference}}}.
\end{aligned}
\end{equation}
Then, the transmitted communication symbol vector ${\bf{d}}_i^{{k_2}}$ can be detected. We use ${\hat{\bf{d}}}_i^{{k_2}}$ to denote the detected symbol corresponding to ${\bf{d}}_i^{{k_2}}$. 
}

{\color{black} Compared with the conventional OFDM JCS system, the CDM gain of the proposed CD-OFDM JCS system can suppress the noise and multiple access interference (MAI). Note that MAI in the JCS scenario includes the co-channel communication interference and the radar echo signals occupying the same frequency band. The CDM gain of the CD-OFDM JCS communication receiver of MTC user 1 is expressed as \cite{Rodger2014principles}
\begin{equation} \label{equ: CDM_gain}
\eta _{C} = \frac{N_c}{NC_2}.
\end{equation}
}
\subsection{The CD-OFDM Radar Signal Processing}
\label{sec:the-acquisition-of-radar-echo}
After obtaining the demodulated communication symbols, we can remove the communication signal from the received superposed signal ${\bf{y}}_{i}^{mc}$, then the $i$th radar echo signal is acquired and can be given by
{\color{black}
\begin{equation} \label{equ: radar signal acquisition}
\begin{aligned}
{\bar{\bf{ d}}_{RX,i}^{k_1}} &=  {\bf{y}}_{i}^{mc} - {{\bar{\textbf{H}}}_{C,21,i}}\frac{\sqrt{P_2}{{\bf{C}}_i^{{k_2}}{ \hat{\bf{d}}  }_i^{{k_2}}}}{{\sqrt {N{C_2}} }} \\
&= \underbrace{ {{\bar{\textbf{H}}}_{R,12,i}}\frac{\sqrt{P_1} {{\bf{C}}_i^{{k_1}}{\bf{d}}_i^{{k_1}}}}{{\sqrt {N{C_1}}}} }_{{\mathop{\rm radar \; echo}}} + \underbrace{ {{\bar{\textbf{H}}}_{C,21,i}}\frac{\sqrt{P_2}{{{\bf{C}}_i^{k_2} \bar{{\bf e}}_i}}}{{\sqrt {N{C_2}} }} + \overline {\bf{n}}_i }_{{\mathop{\rm error \; and \; IPN}}},
\end{aligned}
\end{equation}
where ${\bar{{\bf e}}}_i = {\bf d}_i^{k_2}- \hat{\bf{d}}_i^{{k_2}}$ is the error of communication decoding. Note that $\overline {\bf{n}}_i$ affects both the communication demodulation and radar sensing, and the communication decoding error will directly cause interference to radar sensing. This shows that the high communication reliability is the prerequisite for high radar detection performance. Because the CD-OFDM JCS processing has CDM gain, the error propagation power of CD-OFDM JCS processing is smaller than OFDM JCS processing. This is demonstrated in Section III-D.

The multicarrier radar detection needs $M_s$ transmitted and received JCS symbols. We assume that all the $M_s$ symbols are demodulated with the same DSSS code books ${\bf{C}}_i^{{k_1}}$ and ${\bf{C}}_i^{{k_2}}$. Then, \eqref{equ: radar signal acquisition} can be stacked into matrix expression as follows:
\begin{equation} \label{equ: radar signal acquisition_matrix}
{\bar{\textbf{D}}}_{RX}^{k_1} = \underbrace{ {{\bar{\textbf{H}}}_{R,12}}\odot\frac{\sqrt{P_1}{{\bf{C}}_i^{{k_1}}{\textbf{D}}^{{k_1}}}}{{\sqrt {N{C_1}}}} }_{{\mathop{\rm radar \; echo}}} + \underbrace{ {{\bar{\textbf{H}}}_{C,21}} \odot \frac{\sqrt{P_2}{{{\bf{C}}_i^{k_2} \bar{\textbf{E}} }}}{{\sqrt {N{C_2}} }} + \overline {\textbf{N}} }_{{\mathop{\rm error \; and \; IPN}}},
\end{equation}
where $\odot$ is the Hadamard product, ${\bar{\textbf{D}}}_{RX}^{k_1}$, ${\textbf{D}}^{{k_1}}$, ${\bar{\textbf{H}}}_{R,12}$, ${\bar{\textbf{H}}}_{C,21}$, $\bar{\textbf{E}}$ and $\overline {\textbf{N}}$ are stacked by $\{ {{\bf{\bar d}}_{RX,i}^{{k_1}}}\}_{i = 0}^{{M_s} - 1}$, $\{ {{{\bf d}}_{i}^{{k_1}}}\}_{i = 0}^{{M_s} - 1}$, $\{\bar{\textbf{h}}_{R,12,i}\}_{i=0}^{M_s -1}$, $\{\bar{\textbf{h}}_{C,21,i}\}_{i=0}^{M_s -1}$, $\{{\bf e}_i\}_{i=0}^{M_s -1}$ and $\{\overline {\bf{n}}_i\}_{i=0}^{M_s -1}$, respectively, with dimension $N_c \times M_s$. Note that ${\textbf{D}}^{{k_1}}$ is both the transmitted JCS communication and the radar sensing reference symbols of user 1, and the sensing processing will not deteriorate the communication performance, because the radar echo signal is the reflection of communication beam that also exists in the conventional CD-OFDM and OFDM individual communication system.

}

{\color{black}
Based on the phase variance between ${\bar{\textbf{D}}}_{RX}^{k_1}$ and ${\bar{\textbf{D}}}_{TX}^{k_1} = \frac{\sqrt{P_1}{{\bf{C}}_i^{{k_1}}{\textbf{D}}^{{k_1}}}}{{\sqrt {N{C_1}}}}$, the Doppler shift and the range between user 1 and user 2 can be detected \cite{Sturm2011Waveform}. Here, the Doppler shift and the time delay of our interest are $f_{d,R,0}$  and $\tau_{R,0}$ expressed in \eqref{equ:f_d_r} and \eqref{equ:tau_Rl}, respectively, which are both parameters of the LoS path. 
The complex-valued element-wise division of ${\bar{\textbf{D}}}_{RX}^{k_1}$ over ${\bar{\textbf{D}}}_{TX}^{k_1}$ can be given by
\begin{equation} \label{equ:echo and transmitting_matrix}
\begin{aligned}
{\bar {\textbf{D}}_{div}} &= \frac{{\bar{\textbf{D}}}_{RX}^{k_1}}{{\bar{\textbf{D}}}_{TX}^{k_1}} \\
&= {\bar{\textbf{H}}}_{R,12} + \frac{\sqrt{(NC_1)} \; \overline {\textbf{N}}}{\sqrt{P_1}{{\bf{C}}_i^{{k_1}}{\textbf{D}}^{{k_1}}}} + \sqrt{\frac{NC_1 P_2}{NC_2 P_1}} \frac{ {{\bar{\textbf{H}}}_{C,21}} \odot {{{\bf{C}}_i^{k_2}\bar{\textbf{E}}}}} {{{\bf{C}}_i^{{k_1}}{\textbf{D}}^{{k_1}}}},
\end{aligned}
\end{equation} 
where ${\bar{\textbf{H}}}_{R,12}$ contains the perfect information of the Doppler and delay, the last two terms are noise and error propagation of SIC, respectively. Here, the division between the frequency-domain symbols is equal to the correlation operation in time-domain, ${\bar {\textbf{D}}_{div}}$ is the correlation matrix between the transmitted JCS signals and the received echo signals, and ${\bar{\textbf{H}}}_{R,12}$ can be expressed by
\begin{equation} \label{equ:H_R12}
{\bar{\textbf{H}}}_{R,12} = {\bf{A}} \odot \left({{{\bf{ k}}}_R} {{{\bf{ k}}}_D^{T}}\right),
\end{equation} 
where ${\bf{A}}$ is the matrix of radar channel fading with the ($m,i$)th entry being the fading of the $m$th subcarrier of the $i$th CD-OFDM symbol, ${{{\bf{ k}}}_R}\left( m \right) = \exp \left( - j2\pi m\Delta f \tau_{R,0} \right)$, and ${{{\bf{ k}}}_D}\left( i  \right) = \exp \left( j2\pi i {T_{s}} f_{d,R,0} \right)$. Note that ${\bar{\textbf{D}}}_{TX}^{k_1}$ must not have entry 0, or else the phase shift matrix ${\bar{\textbf{D}}}_{div}$ does not exist, and then radar detection will be invalid. Here, we prove that $N{C_1}$ and $N{C_2}$ should both be odd to ensure that ${\bar{\textbf{D}}}_{TX}^{k_1}$ does not have entry 0, which is included in \textbf{Theorem} \ref{Theo:2}. 
}

\begin{Theo} \label{Theo:2}
	{\rm 
		The CD-OFDM symbol is $\bar{\bf{d}}_{TX,i}^k = {\bf{C}}_i^k{\bf{d}}_i^k$, where ${\bf{C}}_i^k$  is the Walsh-Hadamard code book with dimension ${N_c} \times NC$, ${N_c}$  is the number of subcarrier, $NC$ is the number of code channel, each entry of ${\bf{C}}_i^k$  is in set $\{  + 1, - 1\} $, and  ${\bf{d}}_i^k$ is the symbol vector modulated with M-ary quadrature amplitude modulation (QAM) constellation. Only when $NC$ is odd, can $\bar{\bf{d}}_{TX,i}^k$ never contain entry 0. 
		\begin{proof}			
			The proof is provided in Appendix \ref{Theo:B}. 
		\end{proof}
	}
\end{Theo}

\begin{figure*}[!ht]
	\normalsize
	\setcounter{equation}{38}
	\begin{equation}\label{equ:error-propagation-CD}
		P_{\bar e} \left( \sigma \right) = E\{ \left| \bar e \right|^2 \} = \sum\limits_{r_1,r_2}^{\sqrt{M}} {p_d}\left( {{s_{{r_1}}}} \right) {p_d}\left( {{s_{{r_2}}}} \right) \times \left[ \sum\limits_{r_1^{'},r_2^{'} = 1}^{\sqrt{M}} {{p_{r_1^{'},{r_1}}} \left(\sigma \right) {p_{r_2^{'},{r_2}}} \left(\sigma \right) \left( {{{\left| {{s_{{r_1}}} - {s_{r_1^{'}}}} \right|}^2} + {{\left| {{s_{{r_2}}} - {s_{r_2^{'}}}} \right|}^2}} \right)} \right],
	\end{equation}
	\hrulefill 
\end{figure*}

{\color{black} Notice that ${{{\bf{ k}}}_R}$ and ${{{\bf{ k}}}_D}$ are FFT and IFFT vectors, respectively. By applying IFFT and FFT to each column and each row of ${\bar{\textbf{D}}_{div}}$, respectively, we can obtain the Doppler shift and time delay at the modulus peaks of the transformed matrix \cite{Sturm2011Waveform}. }

\setcounter{equation}{30}

{\color{black}
The IFFT of the $i$th column of ${{\bar{\textbf{D}}}_{div}}$ and the FFT of the $m$th row of ${\bar{\textbf{D}}_{div}}$ are 
\begin{equation}
	\begin{aligned}
		&{{\bar{\textbf{D}}}_{i,IFFT}}(q) \\
		&\!\!= \!\! \frac{1}{\sqrt{N_c}}\!\!\sum\limits_{m = 0}^{N_c - 1} \!\! {{\bf{ k}}_D}\left( i  \right) {\bf{A}}\left( {m, i} \right){{{\bf{ k}}_R}\left( m \right)\exp \left( {j\frac{{2\pi }}{N_c}q m} \right)} \! + \! {\bar w}_{i,q},
	\end{aligned}
\end{equation}
and
\begin{equation} \label{equ:D_mfft}
	\begin{aligned}
		&{{\bar{\textbf{D}}}_{m,FFT}}(l) \\
		&\!\!= \!\! \frac{1}{\sqrt{M_s}}\!\!\!\! \sum\limits_{i  = 0}^{{M_s} - 1} \!\! {{\bf{ k}}_R}\left( m \right) {\bf{A}}\left( {m, i} \right) {{{\bf{ k}}_D}\left( i  \right)\exp \left( { - j\frac{{2\pi }}{M_s}l i} \right)} \! + \! {\bar w}_{m,l},
	\end{aligned}
\end{equation}
respectively, where $q = 0,1,...,N_c-1$, $l = 0,1,...,M_s-1$, ${\bar w}_{i,q}$ and ${\bar w}_{m,l}$ are both the FFT of the noise and error terms and have the same power as the sum of noise and error. Here, the FFT and IFFT operations are equal to the match-filter.} Let $ind_{d,m}$ and $in{d_{R,i}}$ denote the indices of the modulus peaks of ${\bar{\textbf{D}}}_{m,FFT}$ and ${{\bar{\textbf{D}}}_{i,IFFT}}$ respectively, which can be obtained by one-dimensional search, as given by
\begin{equation} \label{equ:ind_d,m}
in{d_{d,m}} = \left\lfloor {{f_{d,R,0}}T_{s}{M_s}} \right\rfloor, \ in{d_{d,m}} = 0,1,...,M_s-1,
\end{equation}
\begin{equation} \label{equ:ind_R,i}
in{d_{R,i}} = \left\lfloor \tau_{R,0}B \right\rfloor, \ in{d_{R,i}} = 0,1,...,N_c-1,
\end{equation}
where $B = \Delta f \cdot N_c$ is the bandwidth of transceivers.

\renewcommand\arraystretch{1.05}
\begin{table}[h]
	\centering
	\caption{Simulation Parameters}
	\begin{tabular}{m{2cm}<{\centering}|m{1.7cm}<{\centering}|m{3.5cm}<{\centering}}
		\hline
		\hline
		\label{Parameter:simulation}
		{\textbf{Parameter Items}}  & {\textbf{Value}} & {\textbf{Description}}\\
		\hline
		$N_c$ & 1024 & The number of subcarriers\\ 
		\hline
		$M_s$ & 1024 & The number of used CD-OFDM frames per radar detection\\ 
		\hline
		$B$ & 122.88 MHz & The total bandwidth\\
		\hline
		$\Delta f$ & 120 kHz & The subcarrier interval bandwidth\\
		\hline
		
		\color{black} 
		$P_1$ & \color{black} 1 W & \color{black} The transmit power of MTC user 1\\
		\hline
		\color{black} $P_2$ & \color{black} 1 W & \color{black} The transmit power of MTC user 2\\
		\hline
		
		$T$ &  8.33 $\mu s$  & The time of each CD-OFDM symbol\\
		\hline
		$T_g$ & 1.43 $\mu s$ & The guard interval\\
		\hline
		$T_s$& 9.77 $\mu s$ & The time of each CD-OFDM frame\\ 
		\hline
		$f_c$ & 24 GHz & The carrier frequency\\
		\hline
		$M$ & 16 & The number of antenna elements of TxA\\
		\hline
		$N$ & 16 & The number of antenna elements of RxA\\
		\hline
		$L$ & 2 & The number of multipath\\
		\hline
		$R_0$ & 100 m & The distance between two MTC devices\\
		\hline
		$v_{rel}$ & 15 m/s & The relative velocity between two MTC devices\\
		\hline
		$\theta_{R,0}$ & 0 & The AoD and AoA for TxA and RxA of JCS MTC user 1\\
		\hline
		$\theta_{RX,C,0}$ & 0 & The AoD and AoA for TxA and RxA of JCS MTC user 2\\
		
		\hline
		\hline
	\end{tabular}
	\label{Parameter}
\end{table}

By first applying IFFT to each column of ${\bar{\textbf{D}}_{div}}$ and then applying FFT to each row of the IFFT-transformed matrix, radar image matrix is obtained, and the two coordinates of the modulus peak of radar image matrix are $ind_{d,m}$ and $ind_{R,i}$, respectively. 
Based on \eqref{equ:ind_d,m} and \eqref{equ:ind_R,i}, the time delay $\tau_{R,0}$ and the Doppler shift $f_{d,R,0}$ between users 1 and 2 can be obtained. Subsequently, the distance and the radial relative velocity between users 1 and 2 can be respectively expressed as
\begin{equation} \label{equ:distance}
R_{0} = \frac{\tau_{R,0}c_0}{2},
\end{equation}
and
\begin{equation} \label{equ:velocity}
{v_{rel}}  = \frac{f_{d,R,0}{c_0}}{2f_c}.
\end{equation}

\subsection{The OFDM JCS Signal Processing}\label{sec:OFDM_JCS}

The CD-OFDM JCS signal processing method is compatible with OFDM JCS signal processing. 
{\color{black} As shown in the Section \ref{sec:system-model}, \ref{sec:Communication Demodulation and Radar Signal Acquirement of CD-OFDM JCS Signal} and \ref{sec:the-acquisition-of-radar-echo}, our proposed CD-OFDM JCS system achieves CDM gain by precoding the information symbols to a certain number of subcarrier symbols. We can achieve our proposed CD-OFDM JCS system by modifying the baseband signal processing software of OFDM transceivers as demonstrated in the Section \ref{sec:Communication Demodulation and Radar Signal Acquirement of CD-OFDM JCS Signal} and \ref{sec:the-acquisition-of-radar-echo}, without changing the OFDM radio-frequency (RF) front end. Therefore, by modifying the DSSS precoding matrices ${{\bf{C}}_i^{{k_1}}}$ and ${{\bf{C}}_i^{{k_2}}}$ to be identity matrices, our CD-OFDM JCS system can be equivalent to the OFDM JCS system. The detailed software modification that can make CD-OFDM JCS system be compatible with OFDM JCS system is as follows.
	
Substitute both ${{\bf{C}}_i^{{k_1}}}$ and ${{\bf{C}}_i^{{k_2}}}$ with identity matrix, i.e. $\textbf{I}_{N_c}$, make ${\bf{d}}_i^{{k_1}}$ and ${\bf{d}}_i^{{k_2}}$ both be information vectors with dimension $N_c \times 1$, and set both $\frac{1}{{\sqrt {N{C_1}}}}$ and $\frac{1}{{\sqrt {N{C_2}}}}$ to be 1, then, the OFDM JCS signal processing method is the same as the processing procedures in Sections \ref{sec:Communication Demodulation and Radar Signal Acquirement of CD-OFDM JCS Signal} and \ref{sec:the-acquisition-of-radar-echo}. DSSS switch module shown in Fig.~\ref{fig:JSC_signal processing} determines whether CD-OFDM or OFDM JCS processing method is used.

Let $\gamma_{OF}$ denote the SINR of OFDM JCS processing at each subcarrier symbol received by MTC user 1, then the SINR of CD-OFDM JCS processing $\gamma_{CD}$ at the same subcarrier symbol is given by
\setcounter{equation}{36}
\begin{equation} \label{equ: gamma_CD-OFDM}
\gamma_{CD} = \eta_{C}\gamma_{OF},
\end{equation}
where $\eta_{C}$ is the CDM given in \eqref{equ: CDM_gain}.
}

\subsection{{\color{black} Error Propagation Analysis}} \label{sec:error_propagation}

{\color{black}
As demonstrated in \eqref{equ: radar signal acquisition_matrix} and \eqref{equ:echo and transmitting_matrix}, the error propagation term $\bar{\textbf{E}}$ affects the detection accuracy. In this section, we analyze the error propagation performance of CD-OFDM and the compatible OFDM JCS signal processing. 

Because all the entries of $\bar{\textbf{E}}$ in \eqref{equ: radar signal acquisition_matrix} are i.i.d. variables, we use $\bar e$ to denote the typical entry of $\bar{\textbf{E}}$. All the entries of $\frac{{{{\bf{C}}_i^{k_2} \bar{\textbf{E}}}}}{{\sqrt {N{C_2}}}}$ have the same distribution as $\bar e$, as demonstrated in Section~\ref{sec:Signal model}. The expectation of $\left| \bar e \right|^2$ is our interest, i.e., $E\{ \left| \bar e \right|^2 \}$.

The $M$-QAM constellation is equal to the combination of two orthogonal and independent pulse amplitude modulation (PAM), i.e., cophase and quadrature PAM, where each PAM constellation has $\sqrt{M}$ symbols with the same power. We use $a$ to denote the symbol interval of the PAM constellation, $\mathbb{S}_{PAM} = \{s_1,s_2,...,s_{\sqrt{M}}\}$ to denote the PAM symbols set, and $d_c, d_q \in \mathbb{S}_{PAM}$ to denote the cophase and quadrature components, respectively. Here, we use the normalized $M$-QAM constellation with unit power to analyze the error propagation power, therefore, $E\left[ {{d_c}^2} \right] = E\left[ {{d_q}^2} \right] = 0.5$.

If the transmitted code is $s_r$, and the subscript $r = 1,...,\sqrt{M}$, then the received symbol follows a Gaussian distribution $N(s_r, \sigma _{}^{^2})$, where $\sigma^2$ is the variance. Apply the maximum likelihood detection, then the probability that the corresponding decoded symbol falls into $s_{r'}$ is \cite{Rodger2014principles}
\begin{equation}\label{equ:Prob_decode}
p_{r',r} \left(\sigma\right) {\rm{ = }} \!
\begin{cases}
1 - Q\left( {\frac{{{s_{r'}} - {s_r} + a/2}}{{\sigma}}} \right)\! &,\; r'=1\\
Q\left( {\frac{{{s_{r'}} - {s_r} - a/2}}{{\sigma _{}}}} \right)& , \; r'=\sqrt{M}\\
Q\left( {\frac{{{s_{r'}} - {s_r} - a/2}}{\sigma _{}}} \right) \!\! - \! Q\left( {\frac{{{s_{r'}} - {s_r} + a/2}}{\sigma _{}}} \right) \! \! \! \! \! &, \;  else
\end{cases},
\end{equation}
where $Q\left(  \cdot  \right)$ is the Q-function, and $r'=1$ and $r'=\sqrt{M}$ are the situations where the decoded symbols are the marginal PAM constellation symbols. When $r' = r$, the decoding is successful, or else the error power is $ \left| \bar e \right|^2 = {\left| s_r - s_{r'} \right|}^{2}$.

\begin{figure*}[!ht]
	\centering
	\subfigure[BER simulation results of the 1 code channel CD-OFDM JCS processing versus the OFDM JCS processing]{\includegraphics[width=0.327\textheight]
		{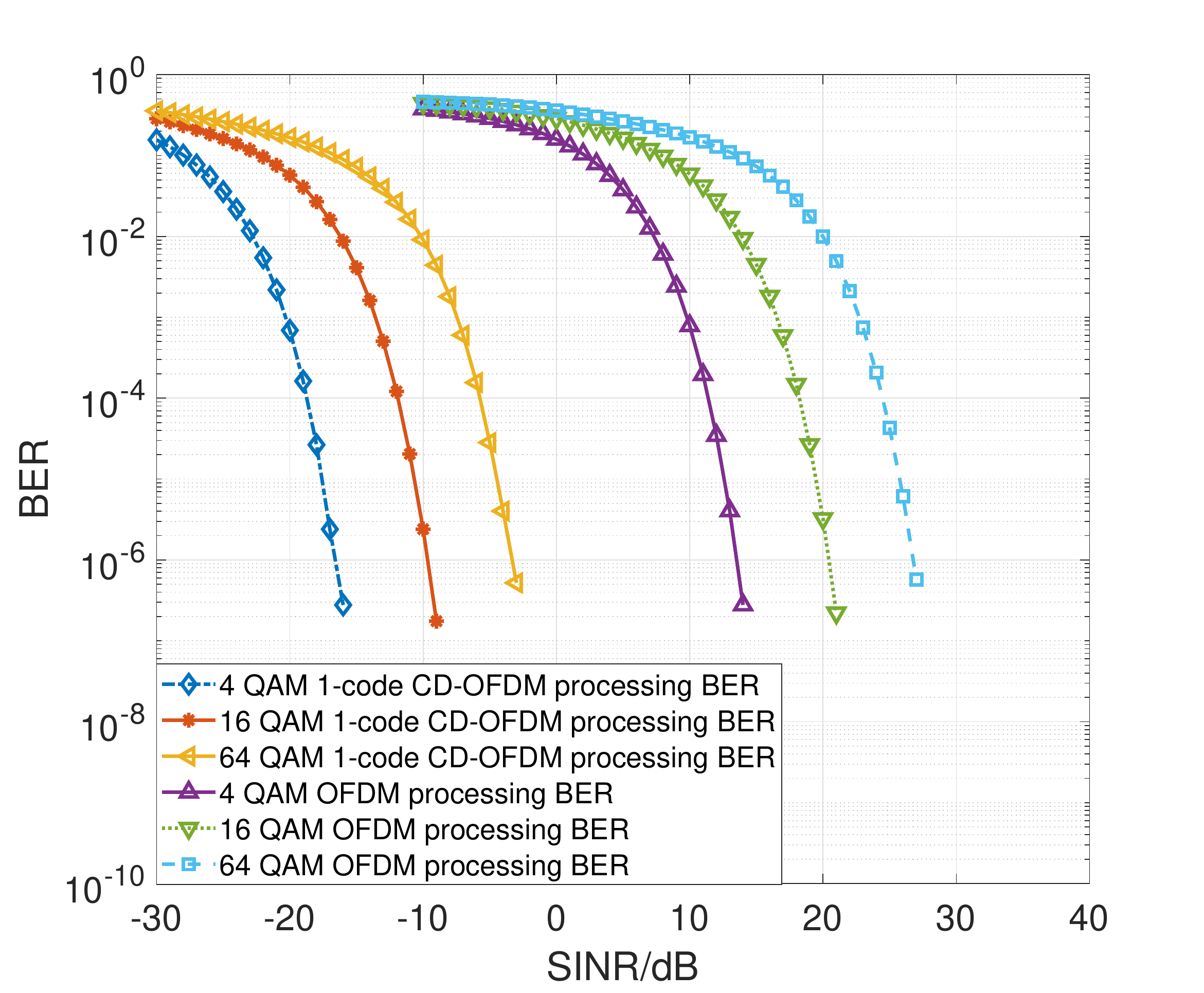}
		\label{figs:BER_1-Noncode}
	}
	\subfigure[BER simulation results of the 255 code channel CD-OFDM JCS processing versus the OFDM JCS processing]{\includegraphics[width=0.34\textheight]
		{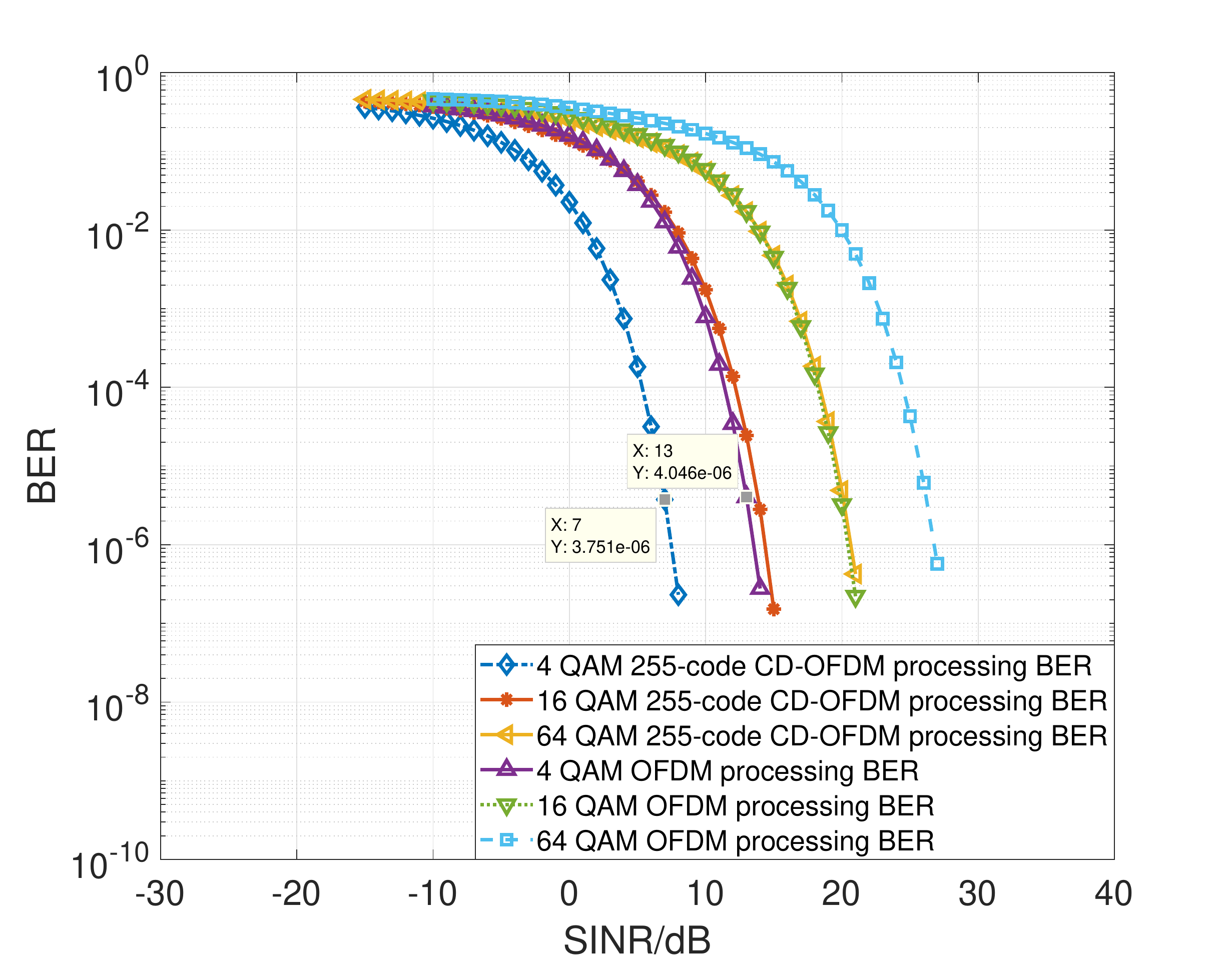}
		\label{figs:BER_255-Noncode}
	}
	\\
	\subfigure[BER simulation results of the 511 code channel JCS CD-OFDM processing versus the OFDM JCS processing]{\includegraphics[width=0.35\textheight]
		{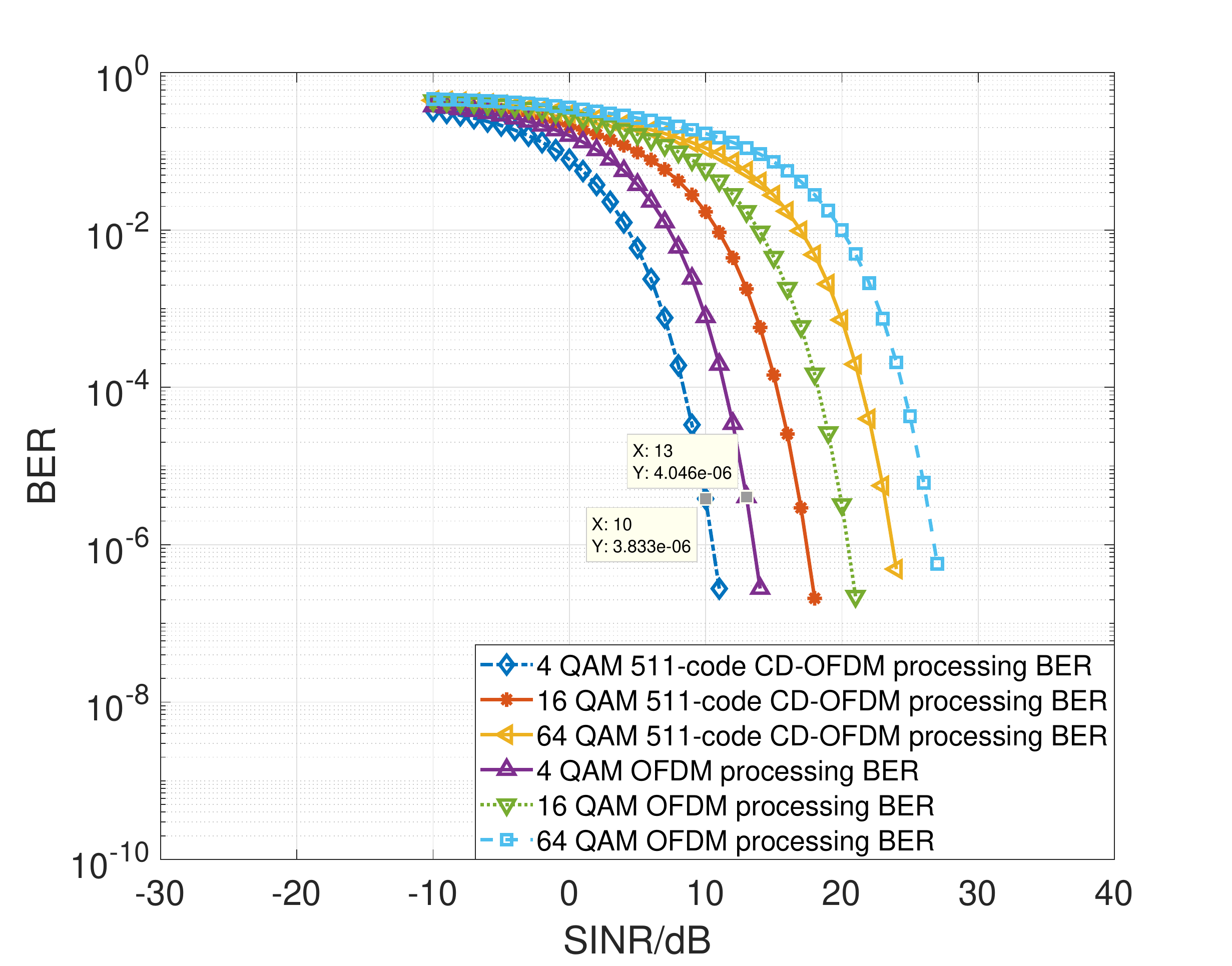}
		\label{figs:BER_511-Noncode} 
	}
	\subfigure[BER simulation results of the TDD OFDM JCS processing versus the OFDM JCS processing]{\includegraphics[width=0.34\textheight]
		{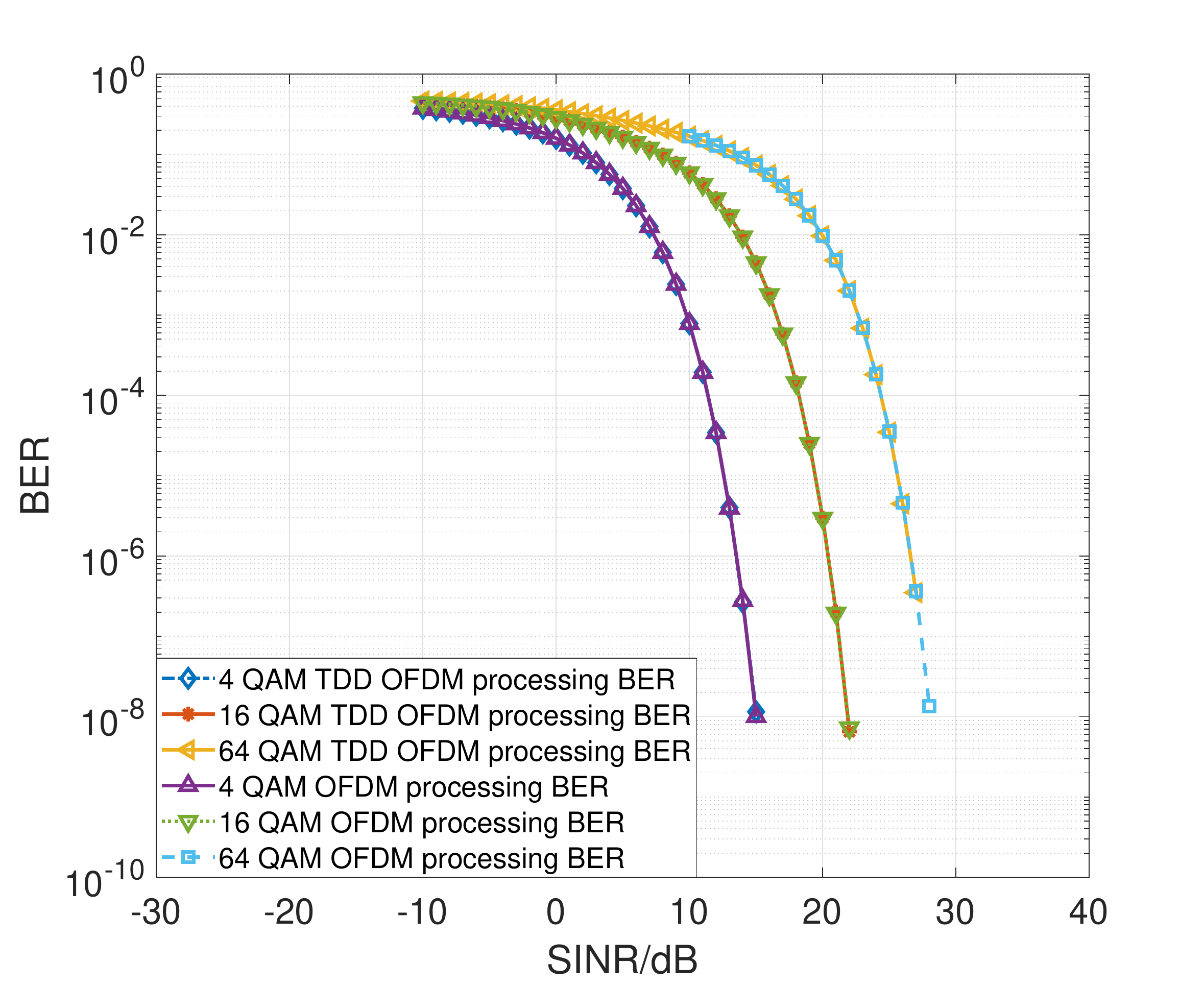}
		\label{figs:TDD_noncode}
	}\\
	\caption{BER simulation results changing against communication SINR.}
	\label{fig:BER}
	
\end{figure*}

Typically, $d_c$ and $d_q$ are i.i.d. variables that follow the discrete uniform distribution. We use ${p_d}(\cdot)$ to denote the probability density function of $d_c$ and $d_q$. We assume $d_c = s_{r_1}$ and $d_q = s_{r_2}$, and the corresponding decoded symbols are ${\hat d}_c = s_{r_1^{'}}$ and ${\hat d}_q = s_{r_2^{'}}$, respectively. Then, the average error propagation power of CD-OFDM JCS system can be given by \eqref{equ:error-propagation-CD}, which is obtained by going through all the combinations of transmitting symbols and decoding symbols.

The IPN variance of the CD-OFDM and OFDM JCS signal processing is $\sigma_{CD}^2 = E\left[ {{d_c}^2} \right]/{\gamma_{CD}}$ and $\sigma_{OF}^2 = E\left[ {{d_c}^2} \right]/{\gamma_{OF}}$, respectively. Then, the average error propagation power of CD-OFDM and OFDM JCS processing is $P_{\bar e}{\left( \sigma_{CD} \right)}$ and $P_{\bar e}{\left( \sigma_{OF} \right)}$, respectively. Here, $P_{\bar e}{\left( \sigma_{CD} \right)}$ and $P_{\bar e}{\left( \sigma_{OF} \right)}$ are both expressed in \eqref{equ:error-propagation-CD}. Because the CD-OFDM JCS processing enjoys the CDM gain, the IPN variance of the CD-OFDM JCS processing is smaller than the OFDM JCS processing, which results in the smaller bit error rate. Thus, the CD-OFDM JCS processing can suppress the error propagation compared with the OFDM JCS processing.
}

\section{{\color{black} Simulation and Numerical Results}} \label{sec:Numerical-results}

We consider a basic situation where two MTC JCS devices move in the same straight direction to demonstrate the performance of the CD-OFDM JCS system. The simulation parameters are listed in TABLE \ref{Parameter:simulation}. Assume that the antenna spacing of RxA and TxA is half wavelength. Based on \eqref{equ:Communication_channel_fre} and \eqref{equ:Radar_channel_Tx_fre}, the channel state matrices for communication and radar sensing are generated at first. Then, JCS signal processing is conducted as shown in Section \ref{sec: Communication}. 

\begin{figure}[!t]
	\centering
	\includegraphics[width=0.47\textwidth]{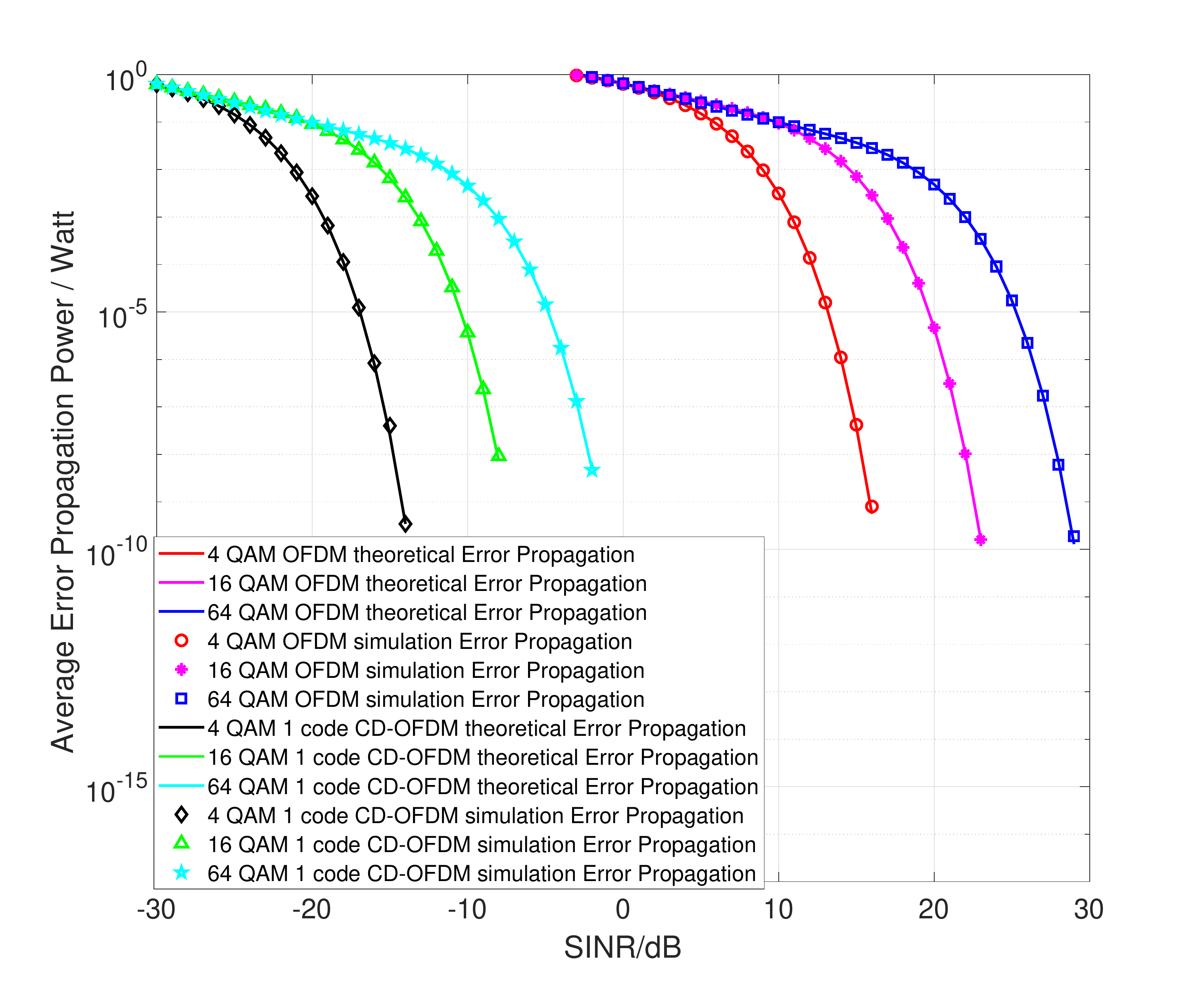}%
	\DeclareGraphicsExtensions.
	\caption{Normalized AEPP of the 1-code CD-OFDM and OFDM JCS processing under 4, 16 and 64 QAM.}
	\label{figs:Error_propagation_Noncode_CD1}
\end{figure}

\begin{figure}[!t]
	\centering
	\includegraphics[width=0.4722\textwidth]{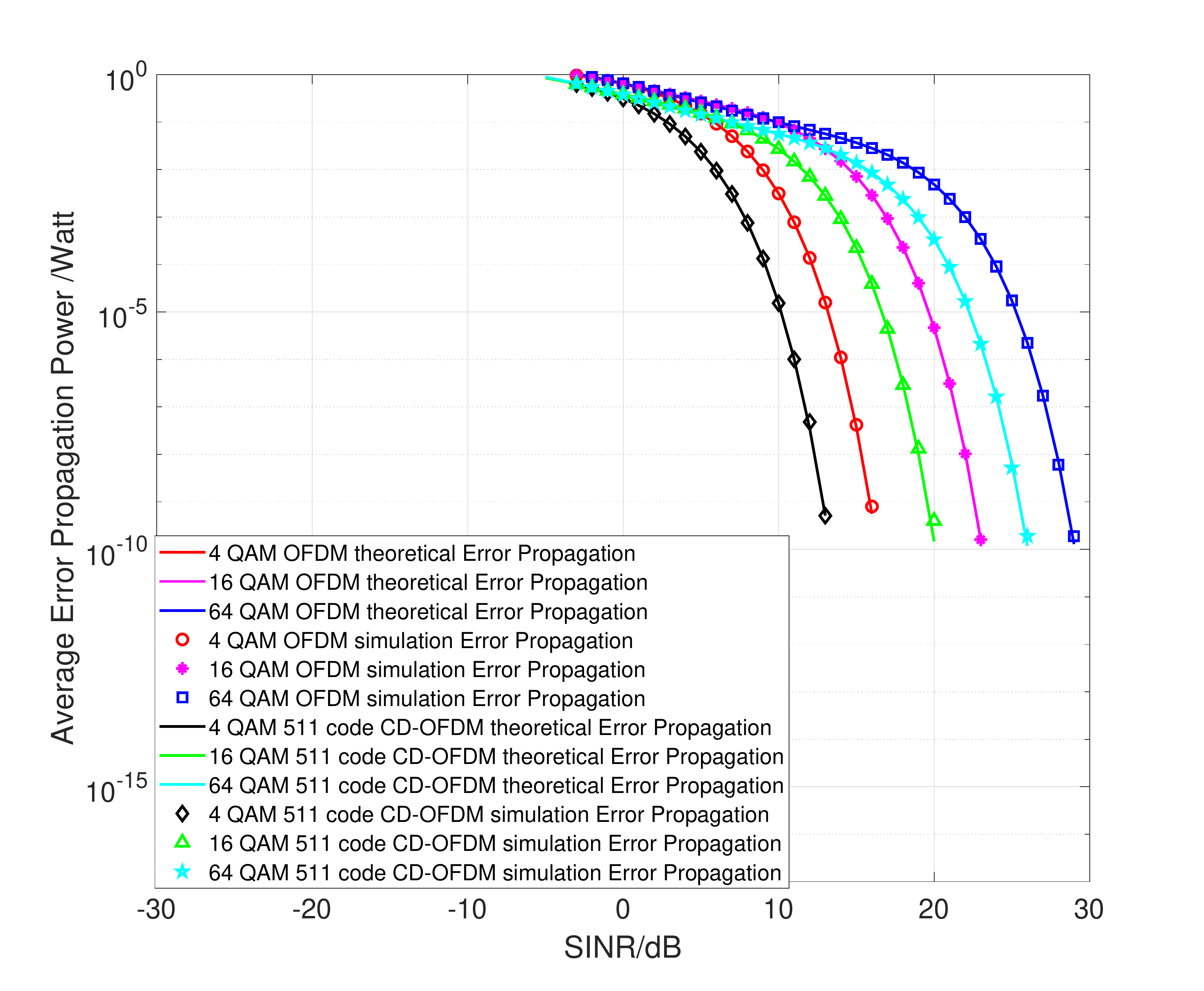}%
	\DeclareGraphicsExtensions.
	\caption{Normalized AEPP of the 511-code and OFDM JCS processing under 4, 16 and 64 QAM.}
	\label{figs:Error_propagation_Noncode_CD511}
\end{figure}


{\color{black} In this section, we present the simulation and numerical results of the JCS performance of the proposed CD-OFDM JCS system and the OFDM JCS system. Because the proposed CD-OFDM JCS system can accommodate the OFDM JCS processing as shown in the Section III-C, the performance curves of ``CD-OFDM and OFDM JCS processing'' in this section can both be achieved by the proposed CD-OFDM JCS system. As a contrast, we introduce the TDD OFDM JCS system from~\cite{Zhang2019JCRS} and present the JCS simulation performance of it, where there is only one JCS device transmitting JCS signal at an assigned time slot, and the mutual interference between communication and radar echo signals is avoided. We adopt M-ary QAM with Gray code as the constellation mapping method. 
}

{\color{black} Fig.~\ref{fig:BER} presents the BER simulation results of the proposed CD-OFDM and the conventional OFDM JCS systems, which verifies our analysis in the Section~\ref{sec:Communication Demodulation and Radar Signal Acquirement of CD-OFDM JCS Signal}.} From Figs.~\ref{figs:BER_1-Noncode}, \ref{figs:BER_255-Noncode} and \ref{figs:BER_511-Noncode}, {\color{black} we can see that the BER of the CD-OFDM JCS processing is lower than the OFDM JCS processing because the MAI interference imposed on the JCS communication demodulation is suppressed. We can identify that when the numbers of code channel are 1, 255 and 511, the values of CDM gain are approximately $10\times\log_{10}(1024)\approx{30.1}$ dB, $10\times\log_{10}(\frac{1024}{255})\approx{6.03}$ dB and $10\times\log_{10}(\frac{1024}{511})\approx{3.02}$ dB according to~\eqref{equ: gamma_CD-OFDM},  respectively.} From Fig.~\ref{figs:TDD_noncode}, we can see that the BER of OFDM JCS processing and TDD OFDM JCS processing is approximate. {\color{black} This is because the interference imposed on the communication demodulation by radar signal is much smaller than the constellation spacing. In a nutshell, the flexible configuration of code channel number leads to the configurable CDM gain of CD-OFDM JCS processing, which makes CD-OFDM JCS system more reliable than the OFDM JCS system.} The JCS MTC device can configure CDM gain to adjust the BER to satisfy the reliability constraint. 

\begin{figure}[!t]
	\centering
	\includegraphics[width=0.485\textwidth]{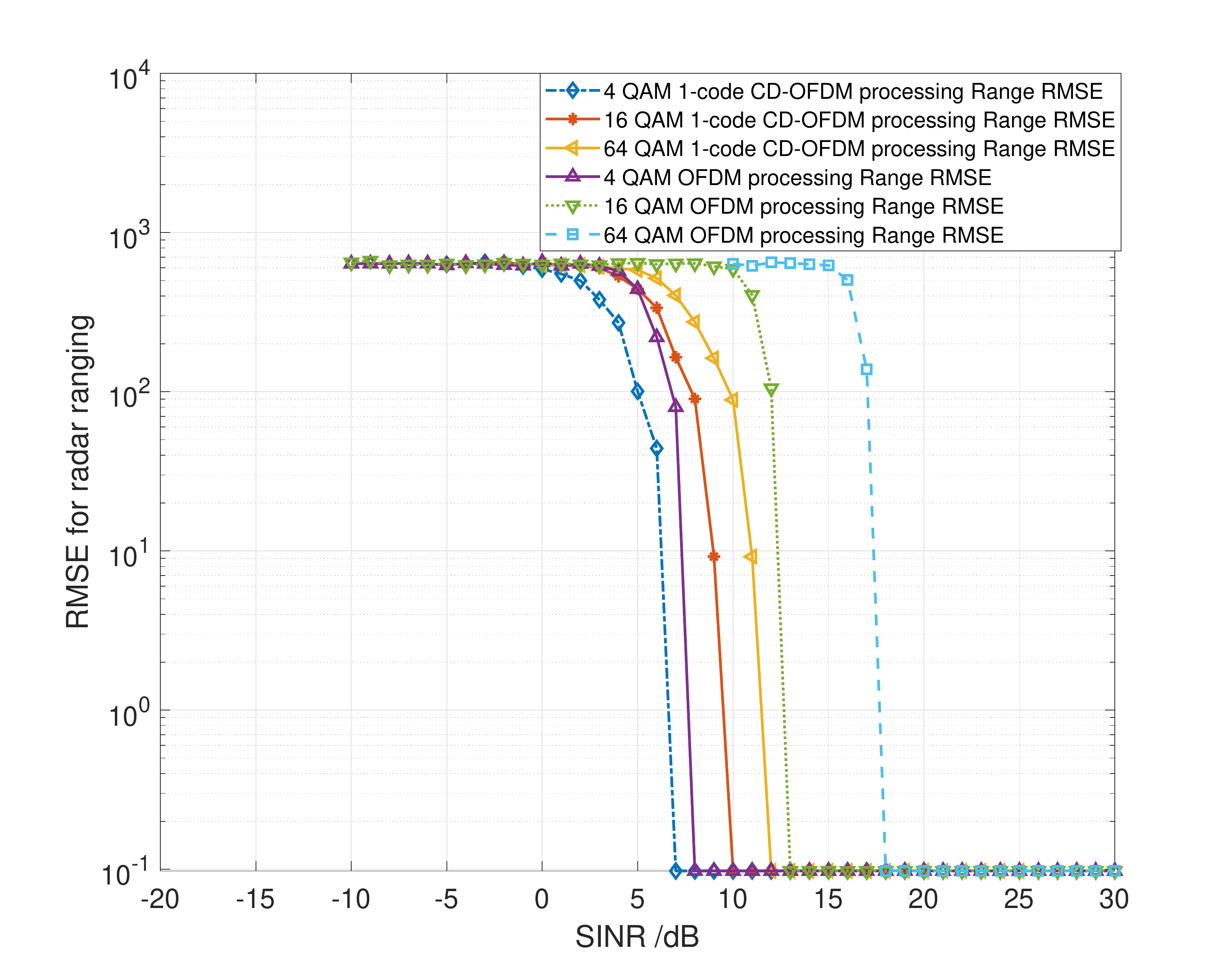}%
	\DeclareGraphicsExtensions.
	\caption{RMSE for radar ranging of the 1 code channel CD-OFDM JCS processing versus the OFDM JCS processing}
	\label{figs:RMSE_range_1_Noncode}
\end{figure}

\begin{figure}[!t]
	\centering
	\includegraphics[width=0.485\textwidth]{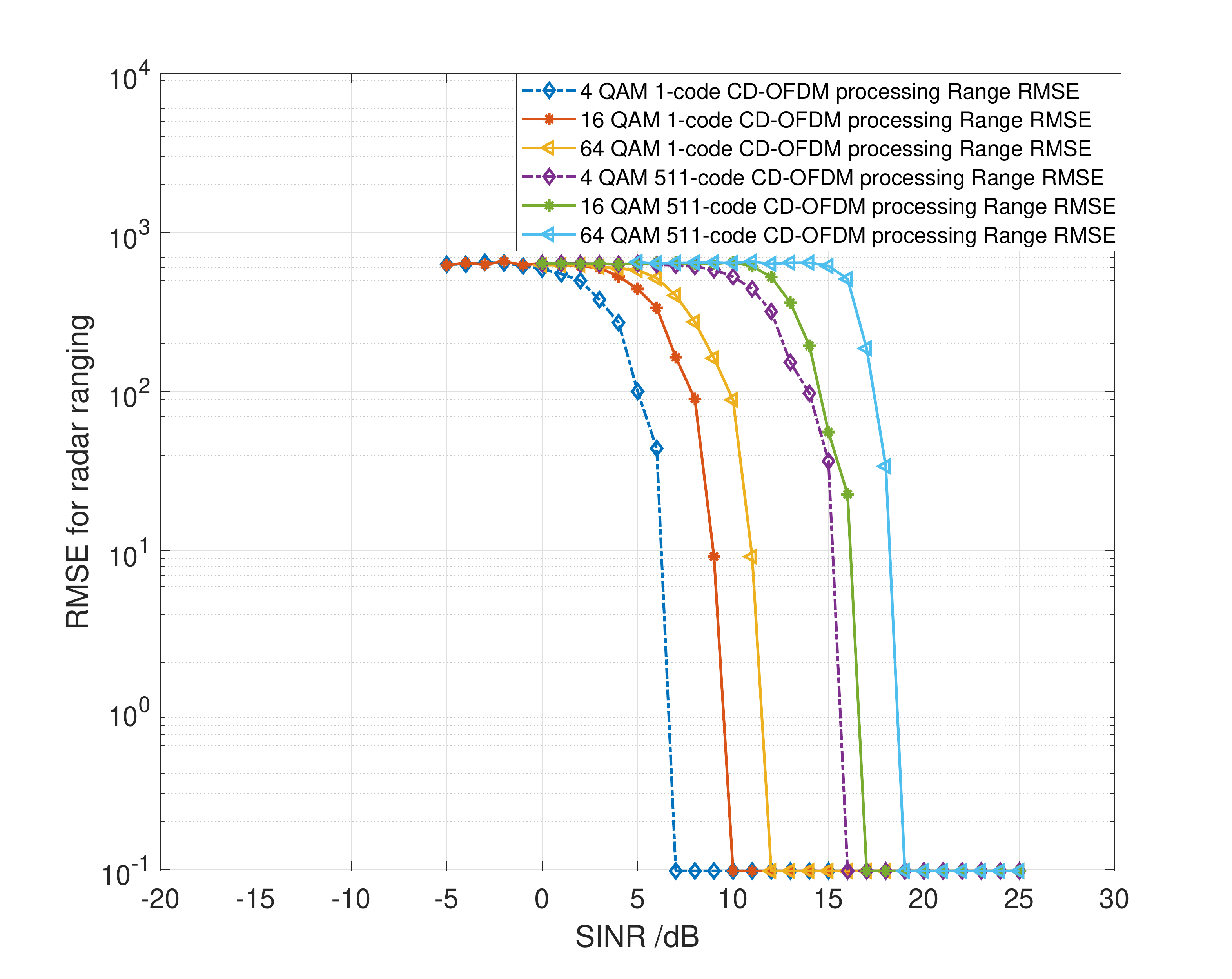}%
	\DeclareGraphicsExtensions.
	\caption{RMSE for radar ranging of the 1 code channel versus the 511 code channel CD-OFDM JCS processing}
	\label{figs:RMSE_range_1_511}
\end{figure}

\begin{figure}[!t]
	\centering
	\includegraphics[width=0.485\textwidth]{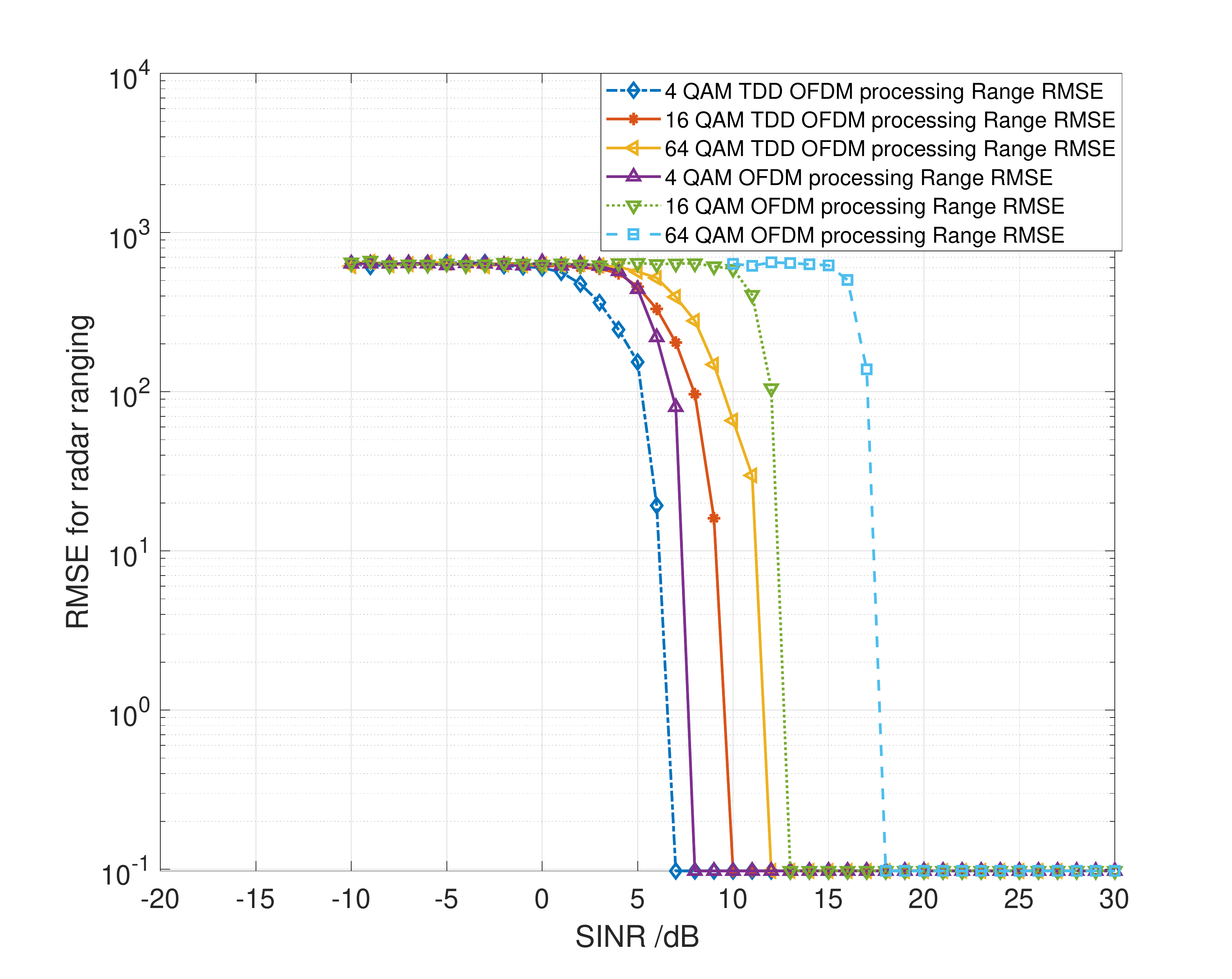}%
	\DeclareGraphicsExtensions.
	\caption{RMSE for radar ranging of the OFDM JCS processing versus the TDD OFDM JCS processing}
	\label{figs:RMSE_range_noncode_TDD}
\end{figure}

\begin{figure}[!t]
	\centering
	\includegraphics[width=0.485\textwidth]{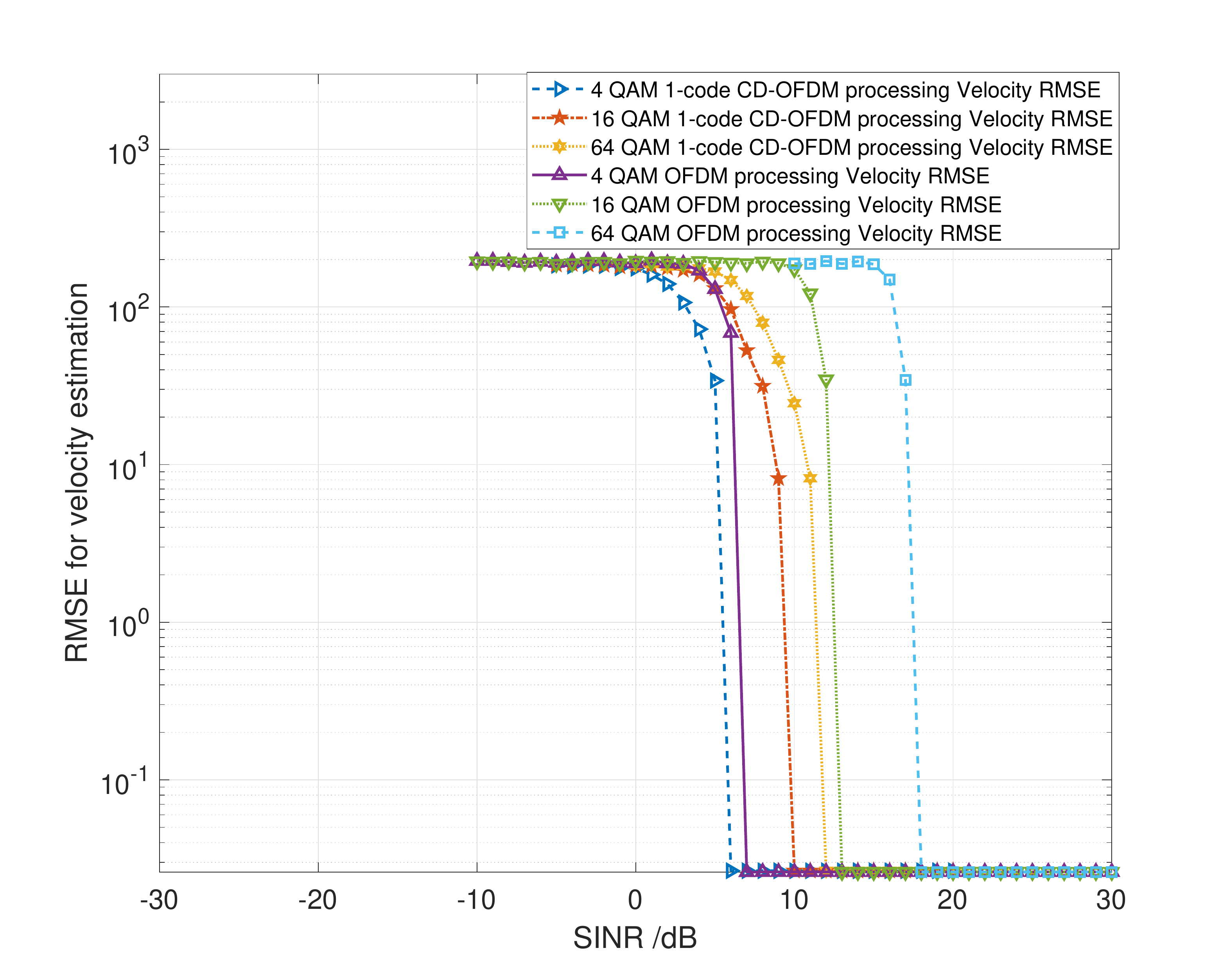}%
	\DeclareGraphicsExtensions.
	\caption{RMSE for velocity estimation of the 1 code channel CD-OFDM JCS processing versus the OFDM JCS processing}
	\label{figs:RMSE_vel_1_Noncode}
\end{figure}

\begin{figure}[!t]
	\centering
	\includegraphics[width=0.485\textwidth]{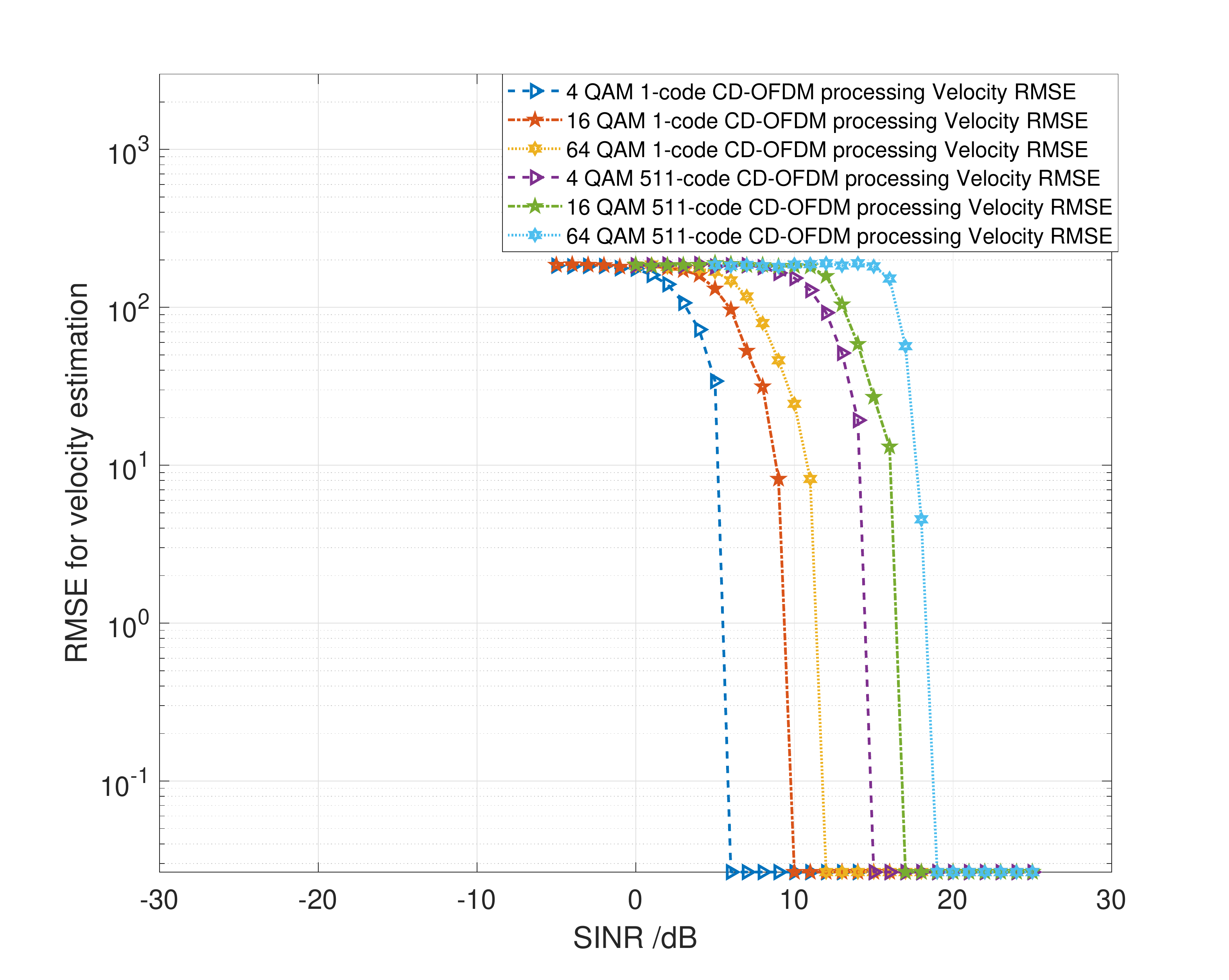}%
	\DeclareGraphicsExtensions.
	\caption{RMSE for velocity estimation of the 1 code channel versus the 511 code channel CD-OFDM JCS processing}
	\label{figs:RMSE_vel_1_511}
\end{figure}

\begin{figure}[!t]
	\centering
	\includegraphics[width=0.485\textwidth]{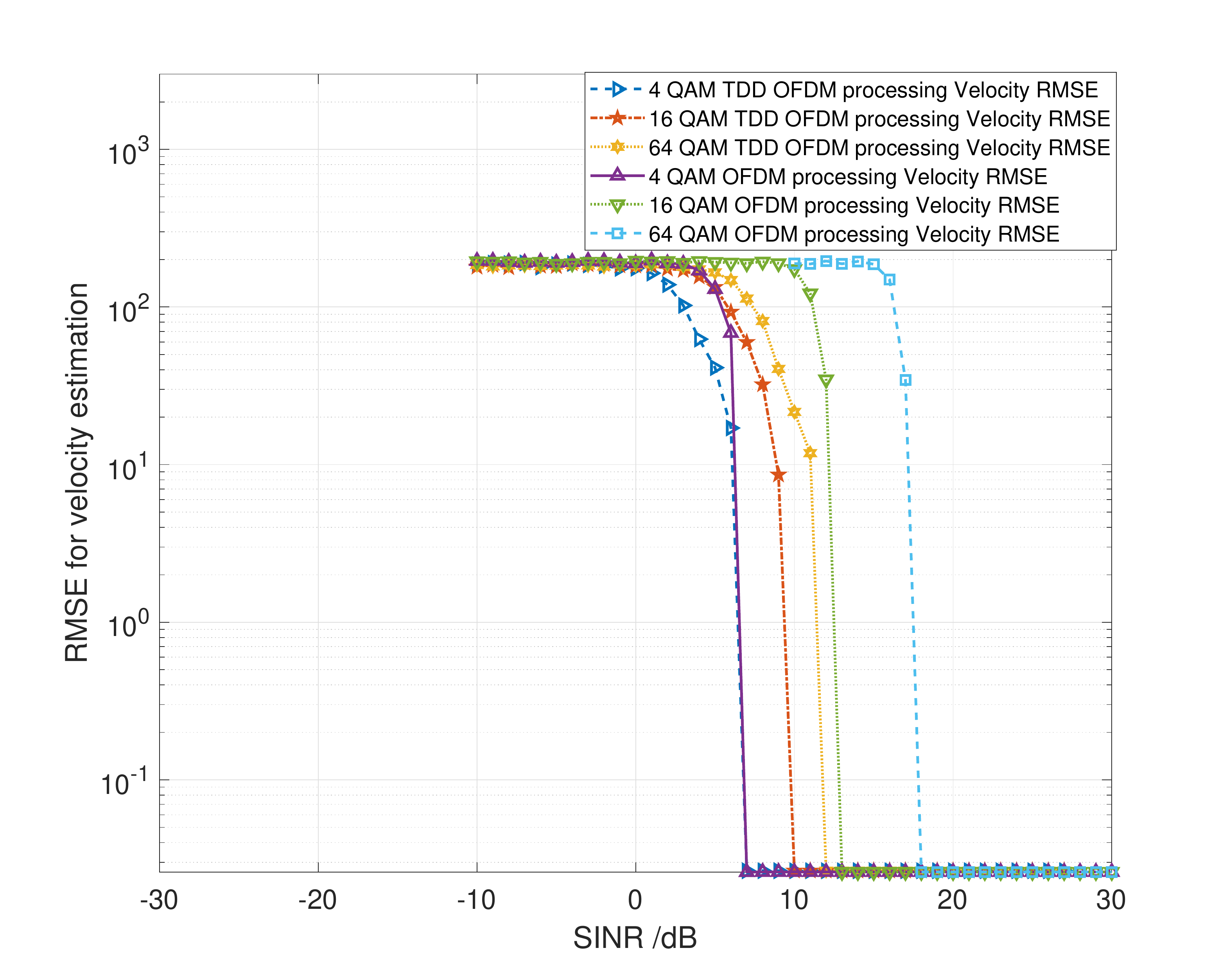}%
	\DeclareGraphicsExtensions.
	\caption{RMSE for velocity estimation of the OFDM JCS processing versus the TDD OFDM JCS processing}
	\label{figs:RMSE_vel_noncode_TDD}
\end{figure}

{\color{black}
	Figs.~\ref{figs:Error_propagation_Noncode_CD1} and \ref{figs:Error_propagation_Noncode_CD511} illustrate the numerical and simulation results of the average error propagation power (AEPP) with normalized unit transmit power, changing against the communication SINR under 4, 16 and 64 QAM. Through the Figs.~\ref{figs:Error_propagation_Noncode_CD1} and \ref{figs:Error_propagation_Noncode_CD511}, we can conclude that the proposed CD-OFDM JCS processing can reduce the AEPP compared with the OFDM JCS processing by providing CDM gain, and the 1 code channel and 511 code channel CD-OFDM JCS processing have 30 dB and 3 dB SINR gain compared with the OFDM JCS processing, respectively, given the same AEPP constraint. This corresponds to the results in \eqref{equ:Prob_decode} and \eqref{equ:error-propagation-CD}. We can also draw the conclusion that the increase of the number of used code channel results in the increase of the AEPP, as the CDM gain decreases according to~\eqref{equ: gamma_CD-OFDM}. 
}

{\color{black}
Figs. \ref{figs:RMSE_range_1_Noncode} to \ref{figs:RMSE_vel_noncode_TDD} present the simulation results of radar ranging and velocity estimation RMSE changing against the communication SINR. Because the frequency-domain IPN in \eqref{equ:echo and transmitting_matrix} is i.i.d., the simulation curves of the velocity estimation in Figs.~\ref{figs:RMSE_vel_1_Noncode}, \ref{figs:RMSE_vel_1_511} and \ref{figs:RMSE_vel_noncode_TDD} have the same variation pattern as the ranging estimation in Figs.~\ref{figs:RMSE_range_1_Noncode}, \ref{figs:RMSE_range_1_511} and \ref{figs:RMSE_range_noncode_TDD}, respectively. As shown in Figs.~\ref{figs:RMSE_range_1_Noncode} and \ref{figs:RMSE_vel_1_Noncode}, the sensing RMSE performance of 1-code channel CD-OFDM JCS processing is better than OFDM JCS processing given SINR, because the CDM gain reduces the AEPP that deteriorates the sensing performance. The increase of code channel number deteriorates the sensing RMSE performance as shown in Figs.~\ref{figs:RMSE_range_1_511} and \ref{figs:RMSE_vel_1_511}. From Figs.~\ref{figs:RMSE_range_1_Noncode} to \ref{figs:RMSE_vel_noncode_TDD}, we can find that the TDD OFDM JCS processing needs smaller SINR than the OFDM JCS processing to guarantee given RMSE constraint, because the TDD OFDM sensing does not risk the error propagation of failed communication demodulation, while the sensing of OFDM JCS system suffers from the error propagation due to SIC-based processing method. In contrast, the CD-OFDM JCS processing can achieve comparable sensing performance to TDD OFDM JCS processing when a small number of code channels are used.}

In the practical operation of MTC JCS system, we have to choose the proper signal processing method to achieve satisfying communication and sensing performance, which is achieved by DSSS switch module as shown in Fig.~\ref{fig:JSC_signal processing}. It is obvious that the CD-OFDM JCS processing obtains the flexible CDM gain, and can satisfy the reliability constraints in the low SINR regime at the cost of occupying more computation resources than the OFDM JCS processing. Besides, the CD-OFDM JCS system can achieve comparable sensing RMSE compared to TDD OFDM JCS system when a proper number of code channels are utilized. 


\section{Conclusion}\label{sec:conclusion}
This work has put forward the CD-OFDM JCS system for low-speed 6G MTC scenarios, which can achieve high communication reliability and sensing performance. The CD-OFDM JCS signal and corresponding SIC-based processing technique have been proposed. Moreover, we have proposed the unified JCS channel model based on MIMO communication and MIMO radar channel models. With the proposed CD-OFDM JCS signal processing method and the unified JCS channel model, we conducted simulation and presented the BER, radar ranging and Doppler estimation RMSE, and average error propagation power of our proposed CD-OFDM JCS system compared with those of the conventional OFDM JCS system. We finally draw the conclusion that the CD-OFDM JCS system can achieve better BER performance than the OFDM JCS system in the low SINR regime at the cost of using more computation resources, while OFDM JCS system can only achieve the same communication performance as CD-OFDM system in the high SINR regime. Moreover, the CD-OFDM JCS system can achieve comparable sensing RMSE performance compared with the TDD OFDM JCS system when a proper number of code channels are used. In practical usage, whether to choose CD-OFDM or OFDM JCS processing mechanism should correspond to the real-time SINR and requirements for sensing and communication of the MTC scenarios. 

\begin{appendices} 

	\section{Proof of Theorem 1} \label{Theo:B}
					The $m$th entry of $\overline {\bf{d}} _{TX,i}^k$ is 
					\begin{equation}
						\overline {\bf{d}} _{TX,i}^k\left( m \right) = \sum\limits_{n = 0}^{NC - 1} {c_{i,m,n}^kd_{i,n}^k},
					\end{equation}
					where $c_{i,m,n}^k$ is the $(m,n)$th entry of  ${\bf{C}}_i^k$,  $m \in \{ {0,...,{N_c} - 1} \}$, $n \in \left\{ {0,1,...,NC - 1} \right\}$, $c_{i,m,n}^k \in \{  + 1, - 1\} $, $d_{i,n}^k$ is the $n$th entry of ${\bf{d}}_i^k$, and $d_{i,n}^k \in {S_{qam}}$ with ${S_{qam}}$ being the set of M-ary QAM constellation points. The constellation points in ${S_{qam}}$ are symmetric, i.e., we can find ${a_2}$ for any ${a_1} \in {S_{qam}}$ to ensure ${a_1} + {a_2} = 0$ where ${a_2} \in {S_{qam}}$. Thus, $\bar d_{i,n}^k = c_{i,m,n}^kd_{i,n}^k$ is in ${S_{qam}}$. Then, $\overline {\bf{d}} _{TX,i}^k$ can be rewritten as
					\begin{equation}
					\overline {\bf{d}} _{TX,i}^k\left( m \right) = \sum\limits_{n = 0}^{NC - 1} {\bar d_{i,n}^k}.
					\end{equation}
					If $NC = 2n$,  $n\in Z$ and $n>0$, then we can find sequences of symmetric constellation points that make $\overline {\bf{d}} _{TX,i}^k\left( m \right)$ be 0. Therefore,  $NC$ cannot be even. On the other hand, if $\sum\limits_{n = 0}^{NC - 1} {\bar d_{i,n}^k}  = 0$ holds for $NC = 2n + 1$, ${n > 0}$, and $n\in Z$, then we can have
					\begin{equation} \label{equ:odd_di}
					\bar d_{i,0}^k = -\sum\limits_{n = 1}^{NC - 1} {\bar d_{i,n}^k}.
					\end{equation}
					{\color{black}
					The right hand side is a sum of even number of constellation points, which can never be an entry of M-ary QAM constellation. Thus, \eqref{equ:odd_di} does not hold. Thus, by reduction to absurdity, $\overline {\bf{d}} _{TX,i}^k\left( m \right)$ can never be 0 if $NC$ is odd.
					}
				
					The proof of \textbf{Theorem} \ref{Theo:2} is completed. 	
\end{appendices}



%

{\small
	\bibliographystyle{IEEEtran}
	\bibliography{reference}
}

%

%
%

\ifCLASSOPTIONcaptionsoff
  \newpage
\fi

\end{document}